\documentclass[11pt]{article}
\pdfoutput=1
\usepackage{jheppub}
\usepackage{amsmath}

\newcommand{\Tr}{\mathrm{Tr}\,}
\newlength{\dummysp}
\newcommand{\half}{\frac{1}{2}}

\newcommand{\beq}{\begin{eqnarray}}
\newcommand{\eeq}{\end{eqnarray}}
\newcommand{\nnn}{ \nonumber \\ }

\newcommand{\chib}{{\bar \chi}}

\newcommand{\vev}[1]{{\langle #1 \rangle}}

\newcommand{\ord}[1]{{{\cal O}(#1)}}
\newcommand{\gappeq}{\mathrel{\rlap {\raise.5ex\hbox{$>$}}
{\lower.5ex\hbox{$\sim$}}}}
\newcommand{\lappeq}{\mathrel{\rlap{\raise.5ex\hbox{$<$}}
{\lower.5ex\hbox{$\sim$}}}}
\newcommand{\myref}[1]{(\ref{#1})}

\newcommand{\ben}{\begin{enumerate}}
\newcommand{\een}{\end{enumerate}}

\newcommand{\hc}{{\rm h.c.}}

\newcommand{\ddd}{\nnn &&}

\newcommand{\bit}{\begin{itemize}}
\newcommand{\eit}{\end{itemize}}

\newcommand{\Ocal}{{\cal O}}

\newcommand{\Abar}{{\bar A}}
\newcommand{\Bbar}{{\bar B}}
\newcommand{\Cbar}{{\bar C}}
\newcommand{\Xbar}{{\bar X}}
\newcommand{\alphabar}{{\bar \alpha}}
\newcommand{\betabar}{{\bar \beta}}

\newcommand{\gammab}{{\bar \gamma}}

\def\[{\left [}
\def\]{\right ]}
\def\({\left (}
\def\){\right )}

\title{On the decoupling of mirror fermions}

\author[a]{Chen Chen,}
\author[a]{Joel Giedt,} 
\author[b]{and Erich Poppitz}

\affiliation[a]{Department of Physics, Applied Physics, and Astronomy, \\ 
Rensselaer Polytechnic Institute, 110 8th St., Troy, New York, 12180, USA}
\affiliation[b]{Department of Physics,   University of Toronto, 
Toronto, ON M5S 1A7, Canada}

\emailAdd{chenc10@rpi.edu}
\emailAdd{giedtj@rpi.edu}
\emailAdd{poppitz@physics.utoronto.ca}

\abstract
{An approach to the formulation of chiral gauge theories on the lattice is to start with a vector-like theory, but decouple one chirality (the ``mirror'' fermions) using strong Yukawa interactions with a chirally coupled ``Higgs" field. While this is an attractive idea, its viability needs to be tested with nonperturbative studies. The model that we study here, the so-called ``3-4-5'' model, is anomaly free and the presence of massless states in the mirror sector is not required by anomaly matching arguments, in contrast to the ``1-0'' model that was studied previously. We have computed the polarization tensor in this theory and find a directional discontinuity that appears to be nonzero in the limit of an infinite lattice, which is equivalent to the continuum limit at fixed physical volume. We show that a similar behavior occurs for the free massless Ginsparg-Wilson fermion, where the polarization tensor is known to have a directional discontinuity in the continuum limit. We thus find support for the conclusion that in the continuum limit of the 3-4-5 model, there are massless charged modes in the mirror sector so that it does not decouple from the light sector. The value of the discontinuity we obtain allows for two  interpretations: either a chiral gauge theory does not emerge and mirror-sector fermions in a chiral anomaly free representation remain massless, or a massless vectorlike mirror fermion appears. We end by discussing some questions for future  study.}
  
\begin{document}
\maketitle
\flushbottom

\section{Introduction}

\subsection{Motivation}

Strongly-coupled chiral gauge theories play an important role
in models of physics beyond the Standard Model. For instance, according
to most attractive channel arguments they may
lead to tumbling dynamics \cite{Raby:1979my}, crucial for
models that generate a hierarchy of scales in
extended technicolor and useful for phenomenological
purposes \cite{Appelquist:1993sg}.  In supersymmetric models,
strong chiral dynamics is prominent in models of dynamical
supersymmetry breaking (see,~e.g.,~the review \cite{Poppitz:1998vd} 
for extensive references),
gauge mediation, and even compositeness
of Standard Model fields as in, for instance,
the ``single sector'' models of Refs.~\cite{ArkaniHamed:1997fq,Luty:1998vr,Terning:1999at}.
Strongly coupled supersymmetric chiral gauge theories
can also lead to novel forms of dark matter
in the hidden sector \cite{Fan:2010is}.\footnote{We only give an illustrative list of particle-physics models using chiral gauge dynamics; for a  more complete list of references, see the Introduction in the recent review article \cite{Poppitz:2010at}.}

Apart
from phenomenological speculations, chiral gauge dynamics
leads to novel and interesting behavior \cite{Shifman:2008cx,Poppitz:2009kz}.  Thus,  it would
be quite interesting to study strongly-coupled chiral gauge theories from first principles.
Techniques such as most attractive channel
arguments \cite{Raby:1979my}, 't Hooft anomaly matching (see, e.g.,
the calculations in \cite{Sannino:2008pz}), scaling
arguments, holomorphy in supersymmetric models,
effective Nambu--Jona-Lasinio--like models, and small volume semiclassical studies  \cite{Shifman:2008cx,Poppitz:2009kz,Poppitz:2009uq} have provided important insight. However, many of the conclusions reached by
these methods need to be checked by a first principles calculation.
To this end, one would like to have a lattice discretization
that yields the correct continuum theory as the lattice
spacing $a$ is sent to zero, so that the successes of
lattice quantum chromodynamics can be carried over
into this new domain.

Vector-like gauge theories are well-defined nonperturbatively with a lattice discretization,
although some fine-tuning of counterterms may be necessary, as for instance in the
case of non-chiral fermions or lattice supersymmetry.  However, 
severe difficulties arise when we try to formulate 
chiral gauge theories on the lattice.  Traditionally,
solving the fermion doubling problem necessitated introducing terms that
explicitly violated chiral symmetry \cite{Nielsen:1980rz,Nielsen:1981hk,Nielsen:1981xu}, 
as in the case of Wilson's fermion
discretization \cite{Karsten:1980wd,Karsten:1981gd}.  However, with the advent of discretizations that
satisfy the Ginsparg-Wilson relation \cite{Ginsparg:1981bj} 
either approximately or exactly---see the work on domain wall fermions \cite{Kaplan:1992bt} and the overlap \cite{Narayanan:1993ss, Narayanan:1994gw}, leading to Neuberger-Dirac operator \cite{Neuberger:1997fp}, recently reviewed in \cite{Kaplan:2009yg}---the fermion-doubling problem is solved while maintaining
a lattice version of chiral symmetry \cite{Luscher:1998pqa}.  Indeed, this striking progress initially led to the
hope that the formulation of lattice chiral gauge theories was at hand.
However, various problems surfaced with this approach.
In the overlap approach, it turned out that the fermion measure
had a gauge-background dependent phase.  These theories also tend to have
a pronounced ``sign problem'' (really, a complex phase problem).
Aspects of these problems have been reviewed in \cite{Golterman:2000hr,Luscher:2000hn,Poppitz:2010at} (there are proposed solutions to some of the problems mentioned above---for instance,  the recent approach of \cite{Golterman:2004qv}, which
involves non-perturbative gauge fixing).

L\"uscher's approach to lattice chiral gauge theories is to define the phase of the fermion determinant
through consistency conditions on the ``measure term" as reviewed in \cite{Luscher:2000hn}. In the case of Abelian chiral gauge theories, the phase can be determined for arbitrary gauge backgrounds and a way of implementing it in practice has been suggested in \cite{Kadoh:2007wz,Kadoh:2007xb}.\footnote{While a numerical implementation of the procedure of \cite{Kadoh:2007wz,Kadoh:2007xb} in 4d might be too expensive, it would be of interest to study the feasibility of implementing it in 2d chiral anomaly free models---such as the 3-4-5 model studied here. As these models are exactly solvable, see \cite{Halliday:1985tg,Kutasov:1994xq}, a numerical investigation  would present a rather nontrivial check on the formalism.} Furthermore,
the nonperturbative solution in the case of nonabelian chiral gauge theories has
yet to be found.  The approach studied in this paper would formulate
such models without having to solve L\"uscher's consistency conditions. We give a brief review of the main ideas in the next Section and recommend \cite{Poppitz:2010at} for an extensive review. 

\subsection{Brief review of    ``mirror decoupling"   with Ginsparg-Wilson fermions}
\label{briefrvw}

The ``single site'' proposal (as originally called in \cite{Bhattacharya:2006dc}) bears some semblance to the
earlier proposals of Refs.~\cite{Eichten:1985ft,Smit:1985nu}. In some
respects the set-up is similar to Wilson quarks, where
doublers are lifted by an irrelevant operator that gives
them an effective $\ord{1/a}$ mass---but at the cost of breaking the chiral symmetry.  The difference here is that only one chirality is supposed to be given a large mass
through a strong interaction with a scalar field.  Historically,
removing chiral fermions from the spectrum was done at the price
of losing gauge invariance \cite{Smit:1980nf,Swift:1984dg,Montvay:1987ys,Montvay:1988av,Aoki:1989xe,Borrelli:1989kw,Hernandez:1995jc}.
For example, in the most recent ``warped domain wall'' proposal  of this kind of Refs.~\cite{Bhattacharya:2005xa,
Bhattacharya:2005mf},
gauge invariance was restored in a particular limit in which the mass of the lightest gauge boson
modes goes to zero.

The mirror-decoupling model proposed in  \cite{Bhattacharya:2006dc} was, instead,
gauge and chirally invariant from the start.  This is the formulation that we study in this paper. In this Section, we only attempt a qualitative description of the intuition underlying the proposal and refer to previous work for the many technical details.

The idea is to begin with a vector-like theory, which is decomposed
into a ``light sector'' of one chirality and a ``mirror sector''
of the opposite chirality.  One then introduces a Yukawa interaction of the mirror fermions 
with a unitary ``Higgs'' field; note that these interactions preserve the gauge symmetry. The purpose of the Yukawa interaction is to lift all states in the mirror
sector up to the cutoff $1/a$ ($a$ is the lattice spacing),
so that only an unbroken chiral gauge theory
survives in the continuum  low-energy theory (the original implementation of \cite{Eichten:1985ft} uses four-fermion interactions, instead of a rapidly fluctuating ``Higgs" field).  However, the lifting of the mirror states is not due to the ``Higgs" obtaining an expectation value: 
instead, the ``Higgs" field, as already alluded to, should be in the disordered (a.k.a. ``strong coupling symmetric") phase---otherwise the gauge symmetry would be broken and one would obtain a broken (massive) gauge theory in the infrared. 

Furthermore, when the Yukawa interaction is taken strong {\it and} the ``Higgs" is in the disordered phase,  strong multi-fermion interactions among the mirror fermions are induced (one can think of integrating out the short-range  fluctuations of the ``Higgs"). Using the lattice strong coupling expansion, it was shown long ago in \cite{Eichten:1985ft} that such interactions {\it can} render all participating fermions massive.  A toy model (that has all the features of \cite{Eichten:1985ft}, but is simpler to study) of decoupling of fermions with strong multi-fermion interactions, showing how all states become gapped in the strong-coupling limit is constructed in Section 3.2  in \cite{Poppitz:2010at}.   

As discussed at length in \cite{Poppitz:2010at},
 the trouble with the earlier proposals of ~\cite{Eichten:1985ft, Smit:1985nu} is that, at the time, a separation of the chiral components of a vector like fermion on the lattice could not be achieved---thus an unambiguous light-mirror separation was absent. Thus all fermions, light and mirror alike, participated in the strong interactions, and, typically, the fermion spectrum was found to be either massive or vector like (see the studies of \cite{Golterman:1992yha} as well as Ref.~\cite{Poppitz:2010at} for more references). In particular, there were no exact chiral symmetries protecting the light fermions from obtaining mass and pairing with the mirror fermions. The situation with respect to lattice chiral symmetries  changed drastically after the advent of the Neuberger-Dirac operator and exact lattice chirality. This allowed the formulation of mirror decoupling with the preservation of exact chiral symmetries and a precise separation between light and mirror fermions \cite{Bhattacharya:2006dc} (we also note the earlier proposal of Ref.~\cite{Creutz:1996xc} along similar lines, but in the domain-wall fermion set up). The precise lattice chiral symmetries allow to formulate precisely on the lattice the 't Hooft anomaly matching conditions. As shown in \cite{Poppitz:2009gt}, these should also be obeyed by strong non-gauge mirror dynamics (in the continuum, these are usually discussed in the strong gauge dynamics framework---as it is difficult to make sense of strong four-Fermi or Yukawa interactions in continuum 4d field theory). 
 
 We stress that the ``mirror decoupling" proposal, both as originally discussed  in \cite{Eichten:1985ft, Smit:1985nu} and as studied  via Ginsparg-Wilson fermions in this paper, relies on non-gauge strong interactions to decouple the mirror fermions. The studies of the spectrum of such models in the past, as well as in this paper, treat the chiral gauge dynamics as a spectator to the mirror-decoupling dynamics. The decoupling dynamics operates at the scale of the lattice spacing---as usual in lattice strong-coupling problems. Including gauge fluctuations is important, of course, but the first task towards implementing the idea is to  (ideally) demonstrate  decoupling of the mirror fermions due to the strong non-gauge dynamics. 

A first step along these lines was taken in \cite{Giedt:2007qg}. The study there was in the context of the two-dimensional
  Schwinger model, which was split into chiral ``light" and ``mirror"  sectors. Strong ``mirror" dynamics was introduced  in an attempt  to decouple one chirality. 
Clearly, such an attempt is destined to fail from the start, because the mirror spectrum is anomalous. However,  the model was chosen because of its minimality, providing an inexpensive tool to study the features of the strong mirror dynamics (for the same reason of cost, the study was restricted to two-dimensional models).
 In the first study of this lattice model~\cite{Giedt:2007qg},
  susceptibilities were measured as a probe of whether or not massless
mirror states were present.  The conclusion of that work was that the susceptibilities
showed no evidence of a massless state in the mirror sector.  However, the
subsequent work \cite{Poppitz:2009gt} used the polarization tensor as a probe,
and found a directional discontinuity at zero momentum, a clear indication of
a massless state.  This agreed with the implications of 't Hooft anomaly
matching (which was shown \cite{Poppitz:2009gt} to hold  for lattice theories with exact chiral symmetries)  that in order to reproduce the mirror-sector anomaly a massless state should
be present.  The interpretation of the two results is that the susceptibilities measured in  \cite{Giedt:2007qg}
did not involve operators that coupled to the massless state, whereas the
polarization tensor did.  It is logical that the polarization tensor would
be a fail-proof probe, since it probes the coupling to all charged states
in the spectrum.  

The motivation of the present work follows exactly from the 't Hooft anomaly
matching argument. As reviewed above, in the theory with an anomalous mirror-fermion content studied in \cite{Giedt:2007qg,Poppitz:2009gt} there were massless mirror-fermion states, as required by 't Hooft anomaly matching. The question we want to address here is whether it is 
 possible that the mirror fermions are lifted by the strong mirror interactions in an
anomaly free theory, where the anomaly matching conditions do not require
the existence of a massless state?  Is the introduction of strong mirror interactions in a manner such that all global mirror chiral 
symmetries are explicitly broken---as envisioned long ago by \cite{Eichten:1985ft}---sufficient to ensure that no massless mirror states are present? Or is there some additional, yet not understood, aspect of the strong mirror dynamics that produces massless mirror states?

Before we continue, let us  summarize some important points about the purpose of this paper that may be a source of concern with the reader. While some of these points were briefly addressed in the discussion above, we think they are important enough to emphasize again:
  \begin{enumerate}
  \item 
The phase intended to make the mirror-sector states composite is a purely lattice phase which does not have a continuum analogue. One can not understand this phase based on intuition coming from continuum gauge theory models. There exist toy models where decoupling in strong-coupling symmetric phases can be shown analytically. There is also a vast literature on strong-coupling symmetric phases in many different lattice theories; for an extensive list of references, see \cite{Poppitz:2010at}.
\item The fact that the gauge interactions are not included in this study may also be a concern. Our purpose here is to investigate if mirror decoupling, due to {\it non-gauge} strong interactions, works in a situation where one hopes it will work: namely when the mirror fermions are in an anomaly-free representation and there are no 't Hooft anomaly matching conditions for global (not gauge) chiral symmetries  that require light mirror states to be present.
\item The reader might also wonder why we do not supplement our numerical study with analytical strong-coupling arguments. The reason this is difficult is as follows: the chirally-symmetric Ginsparg-Wilson Dirac operator used to write mirror actions that explicitly break all unwanted global chiral symmetries of the mirror sector (but preserve the one exact chiral symmetry to be gauged) couples different lattice sites in an exponentially local way. The usual strong-coupling expansion crucially relies on the fact that the interaction does not couple different lattice sites and the spectrum, or the correlation functions, can be determined starting from the single-site theory---as, for example, illustrated in the toy model in \cite{Poppitz:2010at}. The fact that there is an exponentially local coupling also  in the Euclidean time direction makes it also hard to have a Hamiltonian treatment of the strong coupling limit, analytical or numerical.
\end{enumerate}
In the rest of this paper, we will use the polarization
tensor of the mirror sector as a probe to answer the  questions about mirror decoupling  in the two-dimensional ``3-4-5'' model, a theory
with an anomaly free-matter content.\footnote{We should  clearly state that the reason we are studying a two-dimensional Abelian chiral gauge theory is not our intrinsic interest in its dynamics. In fact, it has been understood for a long time, as the ``3-4-5" model can be solved using bosonization \cite{Halliday:1985tg,Kutasov:1994xq}. The solution displays features, which, while interesting, are believed to be related to its  dimensionality. Notably, the massive spectrum is universal and is that of a related vector-like theory---the Schwinger model---while the massless spectrum consists of free composite fermions which saturate  't Hooft anomaly matching for the anomaly-free global symmetries. What motivates using the ``3-4-5" model is the fact that it offers  the simplest arena to study the viability of  lattice formulations of chiral gauge theories: reproducing the solution of the ``3-4-5" model is a test that  any proposed lattice formulation of chiral gauge theories should pass.} 

\subsection{Outline}
The rest of this paper is organized as follows:  

In Section~\ref{descr},
we define the 3-4-5 model---its matter content and action.  In Section~\ref{formal},
we  briefly review some theoretical developments that have appeared
previously  (for a more thorough discussion we refer the reader to the original references \cite{Poppitz:2007tu,Poppitz:2008au,
Poppitz:2008zz,Poppitz:2009gt,Poppitz:2010at}). Various exact (independent of the coupling) identities obeyed by the mirror polarization operator on the lattice were derived there and  provide essential consistency checks
on our computer code.  It will also be seen how anomaly
cancellation proceeds in the lattice model. A separation of
the polarization tensor into one relevant for studying the
mirror sector is derived.  In Section~\ref{formal}, we  describe the
 mirror sector polarization tensor, giving its lengthy
form in Appendix \ref{exppi}.  

Section \ref{numer} details our numerical
studies---both consistency checks of our code  and results for
the behavior of the polarization tensor.  There, we are able
to offer some interpretations in terms of decoupling and we
find disappointing results:  it would appear that in the continuum
limit, the mirror
sector does not decouple.  This is seen through an apparent persistence of
non-analytic behavior in the polarization tensor about $k \to 0$ as we send
the lattice spacing $a$ to zero, by increasing $N=L/a$ with $L$ held fixed,
where $L$ is the linear extent of the lattice in physical units.
Section \ref{concl} summarizes our findings and discusses future
studies that should be performed in order to provide further
evidence for our conclusions.  

In two appendices, we provide
various details that are useful for understanding our work.
In Appendix \ref{exppi} we give the full expression of the
mirror sector polarization tensor.  Finally, in Appendix \ref{fmat}
we describe how the fermion matrix is organized in the basis
that we work, and give its elements explicitly.

\section{The 3-4-5 model}
\label{descr}
The ``3-4-5 model'' is a two-dimensional lattice gauge theory with
U(1) gauge invariance.  It has three Weyl fermion fields, which we denote $A_+$, $B_+$, $C_-$, 
with charge and chirality $3_+$, $4_+$, and $5_-$ respectively.  In addition, there
is a mirror sector, $3_-$, $4_-$, and $5_+$, and we call the respective fields $A_-$, $B_-$, and $C_+$.  In order to
construct all of the Yukawa couplings that are needed, a neutral spectator 
fermion $X_-$, with charge and chirality $0_-$, is also introduced,
together with its mirror sector partner $X_+$, with charge and chirality $0_+$.  Finally, in order to lift the
mirror sector, a unitary ``Higgs'' field $\phi$ with charge $-1$ is introduced.
The 3-4-5 model fields are given  in Table \ref{fc}.

\begin{table}
\begin{center}
\begin{tabular}{|c|c|c|} \hline
Light Field & Mirror Field & Q \\ \hline
$A_+$ & $A_-$ & 3 \\
$B_+$ & $B_-$ & 4 \\
$C_-$ & $C_+$ & 5 \\
$X_-$ & $X_+$ & 0 \\
--- & $\phi$ & -1 \\ 
\hline
\end{tabular}
\end{center}
\caption{Summary of the field content in the 3-4-5 model. \label{fc}}
\end{table}

The dynamics of the gauge field is not supposed to be involved in the
mechanism that lifts the mirror fermions, so in our analysis we will
neglect the gauge field fluctuations and treat it only as a background field.
Thus the action of the ``3-4-5'' model is
\beq 
\label{eq:action}
S &=& S_{\rm light}+S_{\rm mirror}    \nonumber  \\
S_{\rm light} &=& -(\bar{A}_+\cdot D_3\cdot A_+)
-(\bar{B}_+\cdot D_4\cdot B_+)-(\bar{C}_-\cdot D_5\cdot C_-)-(\bar{X}_-\cdot D_0\cdot X_-) \nonumber \\
S_{\rm mirror} &=&S_{\kappa}-(\bar{A}_-\cdot D_3\cdot A_-)
-(\bar{B}_- \cdot D_4\cdot B_-)-(\bar{C}_+ \cdot D_5\cdot C_+)-(\bar{X}_+ \cdot D_0\cdot X_+) \nonumber \\
           &&+S_{\text{Yuk.,Dirac}}+S_{\text{Yuk.,Maj}},
\eeq
Here, $D_q$ is Neuberger's overlap Dirac operator \cite{Neuberger:1997fp}, 
with charge $q$ on the gauge field.  Chiral projections are based on the $\gamma_5$ and
$\hat \gamma_5$ operators, as is typical in the overlap formalism, and described below.
$S_\kappa$ is the gauge invariant kinetic term for the Higgs field, see (\ref{Sk}) below, and $S_{\text{Yuk.,Dirac}}$ and $S_{\text{Yuk.,Maj}}$ are Yukawa interactions
that will be defined shortly.  Our convention is to define:
\beq \label{eq:latticegamma5}
\hat\gamma_{5q} &=& \frac{1}{\sqrt{X_q X_q^\dagger}} X_q \gamma_5 ~, \\
D_q &=& 1-\hat\gamma_{5q} \gamma_5,\nonumber
\eeq
where we used the Wilson kernel ($D_W = -X_q$ with a mass $-M$):
\beq \label{eq:GW}
X_{q,xy} &=& (M-2r) \delta_{xy}
+ \frac{1}{2} \sum_{\mu} \[ (r-\gamma_\mu) \delta_{y,x+\hat\mu} U_\mu^q(x)
+ (r + \gamma_\mu) \delta_{y,x-\hat\mu} U^{q\dagger}_{\mu}(y) \]~,
\eeq
in which $x$, $y$ label two-dimensional lattice sites, $U_\mu(x)=e^{iA_\mu(x)}$, and
$\hat\mu$ is a unit vector in the $\mu$th direction of the lattice.  
In our calculations we make the usual choice $M=r=1$.
Notice that the Wilson kernel is $\gamma_5$ hermitean 
$X_q^\dagger = \gamma_5 X_q \gamma_5$.  Another important property is
that $\hat\gamma_5^2=1$, which enables us to define projection operators.

\begin{figure}
\centering
\includegraphics[width=4in,height=3in]{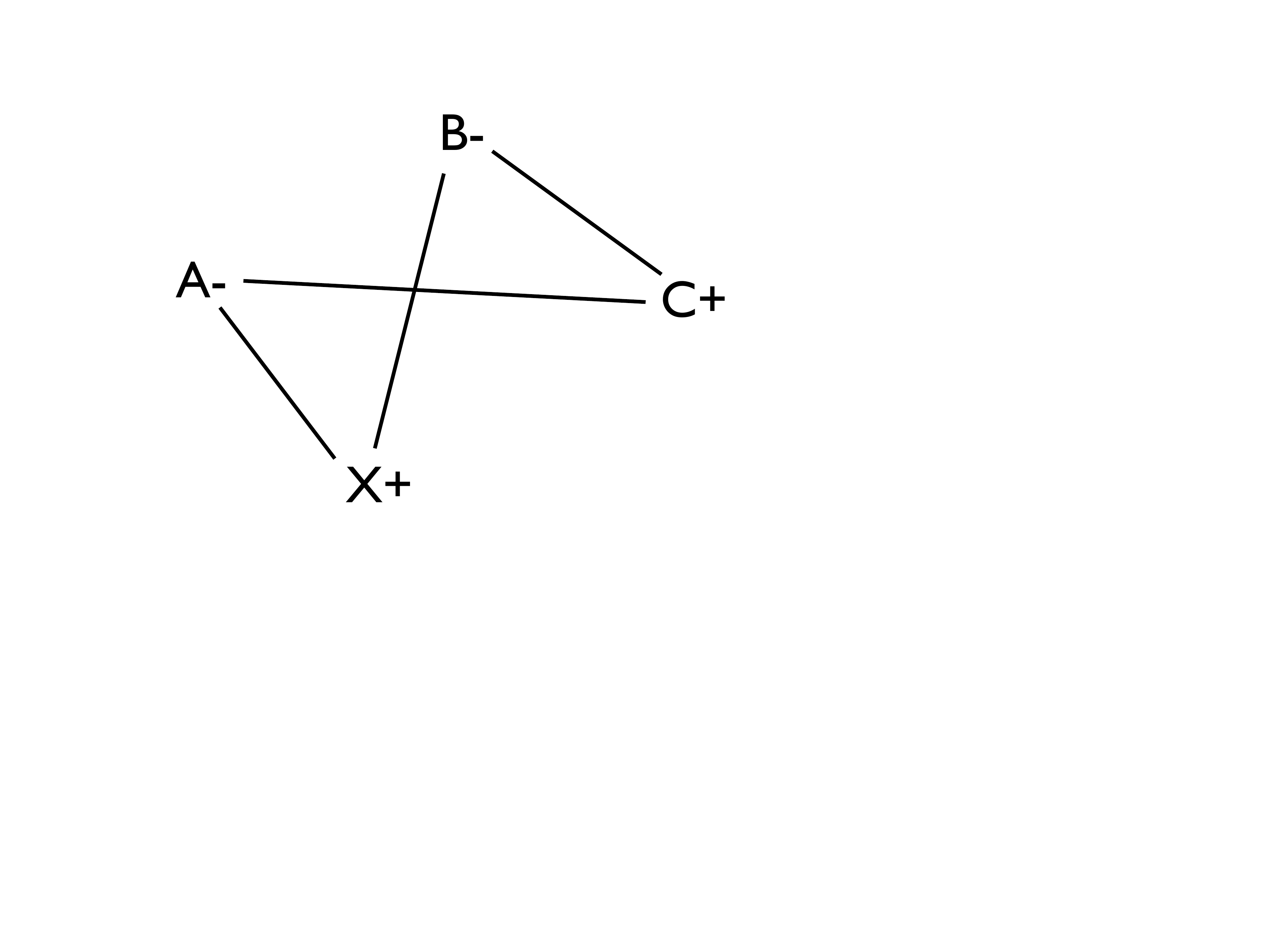}  
\caption{A graph representing the gauge invariant mirror interactions in the 3-4-5 model. The vertices denote the Weyl fermion fields, see Table \ref{fc}, of the mirror theory or their conjugates. The thick connecting lines denote appropriate powers of the unitary ``Higgs" field $\phi$. If a line connects  a fermion field on one vertex with the conjugate of the field on the other vertex, then the corresponding interaction is part of $S_{\text{Yuk.,Dirac}}$. If a line  connects a fermion field with the field (rather than the conjugate) on the other vertex, then the interaction is part of $S_{\text{Yuk.,Maj.}}$.  \label{fig:mirror}}
\end{figure}

The Ginsparg-Wilson fermion that we use in our model allows us to split the 
partition function unambiguously, because of exact lattice 
chiral symmetry, as will be discussed   in the next section. 
The Yukawa interactions of the mirror sector are written following the principles outlined in the caption of Figure \ref{fig:mirror}, see also \cite{Poppitz:2010at}.  Explicitly, they take the following form:
\beq \label{eq:yukawa_interaction} 
S_{\text{Yuk.,Dirac}} &=&
y_{30} \Abar_- X_+ \phi^{-3} + y_{40} \Bbar_- X_+ \phi^{-4} + y_{35} \Abar_- C_+ \phi^{2}
+ y_{45} \Bbar_- C_+ \phi \ddd + y_{30} \bar X_+ A_- \phi^3 + y_{40} \bar X_+ B_- \phi^4
+ y_{35} \bar C_+ A_- \phi^{-2} + y_{45} \bar C_+ B_- \phi^{-1}      \nonumber    \\ 
S_{\text{Yuk.,Maj.}} &=&
h_{30} A_-^T \gamma_2 X_+ \phi^3 + h_{40} B_-^T \gamma_2 X_+ \phi^4 
+ h_{35} A_-^T \gamma_2 C_+ \phi^8 + h_{45} B_-^T \gamma_2 C_+ \phi^9
\ddd - h_{30} \Xbar_+ \gamma_2 \Abar_-^T \phi^{-3} - h_{40} \Xbar_+ \gamma_2 \Bbar_-^T \phi^{-4}
- h_{35} \Cbar_+ \gamma_2 \Abar_-^T \phi^{-8} 
\ddd - h_{45} \Cbar_+ \gamma_2 \Bbar_-^T \phi^{-9}.
\eeq
Each of the  terms in (\ref{eq:yukawa_interaction}) has an implicit sum over lattice sites:  $\Abar_- X_+ \phi^{-3}
= \sum_x \Abar_{-,x} X_{+,x} \phi_x^{-3}$.

In (\ref{eq:yukawa_interaction}), we have included every possible Yukawa term consistent with the $U(1)$ gauge symmetry,
in order to explicitly break all other global symmetries of the mirror sector. 
This is because  any additional global symmetry in the mirror sector which is not explicitly broken
may come with an unwanted anomaly and consequently result in massless excitations at low energy due to 't Hooft anomaly matching. 
Because this is a two-dimensional model, in addition to $S_{\text{Yuk.,Dirac}}$, 
we can  also write down Majorana Yukawa interactions $S_{\text{Yuk.,Maj.}}$, since 
the charge conjugate of a chiral Weyl fermion in two dimensions has
the same chirality (the studies of the anomalous 1-0 model showed that Majorana couplings play an important role).

Finally, as already mentioned, the field $\phi_x=e^{i \eta_x}$, $\left| \eta \right| \leq \pi$, 
is a unitary Higgs field of charge $-1$ with a kinetic term:
\beq
S_\kappa=\frac{\kappa}{2} \sum_x \sum_{\mu}[2-(\phi^*_x U^*_\mu(x) \phi_{x+\hat\mu}+h.c.) ]~.
\label{Sk}
\eeq
At small $\kappa$, the scalar theory is in a symmetric phase (as opposed to
algebraic ordering at large $\kappa$), where the unitary Higgs field has a
correlation length of order the lattice spacing. Heuristically, one imagines that ``integrating out" the short-range $\phi$-fluctuations generates   multifermion interactions among the mirror fields that break all global chiral symmetries.

\section{Review of formal developments}
\label{formal}

\subsection{Splitting of the partition function}
 
As usual in a vector-like theory, the action 
introduced in the last section completely defines the theory via 
the path integral:
\beq
\label{v11}
Z[A] = \int [d\psi d\bar\psi d\phi] e^{S_{\text{light}}[A]+S_{\text{mirror}}[A]},
\eeq
where $[d\psi d\bar\psi]$ is the fermion measure of the vector-like theory,
and represents a Grassmann integral over all of the fermion fields $\psi = A,B,C,X$ (and
the corresponding conjugates).
As mentioned above, we are not going to integrate over the gauge field, since we treat it
as a background field.  
Due to the exact lattice chiral symmetry,
in the case where the light and mirror
sectors are separately anomaly free, the partition function can be split 
into light and mirror parts in a way which is  nonsingular over the entire field space  \cite{Poppitz:2007tu}. 

The splitting of the partition function into light and mirror sectors 
is achieved by decomposing the Dirac fermions $A,B,C,X$ into chiral parts through the
chiral eigenvectors of $\gamma_5$ and $\hat\gamma_5$, as we now briefly review. In our convention,  $\gamma_5$ generates chiral transformations on
unbarred fields, while $\hat\gamma_5$ generates chiral transformations on the barred fields.
 The operator $\hat \gamma_{5}$ depends
on $q$ since it involves $U_\mu^q(x)=e^{iqA_\mu(x)}$; see (\ref{eq:latticegamma5}) and (\ref{eq:GW}) above.
For this reason we need four sets of eigenvectors, corresponding to the four fermions with charges $q=0,3,4,5$.
In the mirror sector, the $\hat\gamma_5$ eigenvectors we need are:
\beq
\label{eq:hatgamma5}
\hat\gamma_{5A} w_{iA} = w_{iA}, \quad \hat\gamma_{5B} w_{iB} = w_{iB}, \quad
\hat\gamma_{5C} u_{iC} = -u_{iC}, \quad \hat\gamma_{5X} u_{iX} = -u_{iX},
\eeq
where we did not explicitly indicate the eigenvector's dependence on the gauge background.
When 
 calculating the polarization tensor, see Section \ref{polarization}, we will differentiate the partition function 
with respect to $A_\mu$ and then set $A_\mu=0$. Thus, when using the tools developed in \cite{Poppitz:2007tu} to calculate the polarization operator, all we will
ever need is the $A_\mu=0$ limit of the eigenvectors (\ref{eq:hatgamma5}).
The $A_\mu = 0$ eigenvectors of   $\gamma_5, \hat\gamma_5$ are:
\beq
\gamma_5 v_i = v_i, \quad \gamma_5 t_i = -t_i \quad
\hat \gamma_5 u_i = -u_i,  \quad \hat\gamma_5 w_i=w_i ~.
\label{eq:gamma5}
\eeq
In the case of the eigenvectors $v_i$ and $t_i$, the index $i$
can be taken to correspond to momenta on the lattice.  
In the case of vanishing gauge field, $A_\mu=0$,
the eigenvectors $u_i$ and $w_i$ of the operator $\hat\gamma_5$
can also be taken to be momentum eigenstates. Explicit expressions in the  convention of this paper are given
in Eq.~(C.2) of \cite{Poppitz:2009gt}.

After expanding the fermion fields in terms of corresponding eigenvectors, 
the partition function (\ref{v11}) splits into light and mirror parts, as in \cite{Poppitz:2007tu}: 
\beq \label{eq:Zsplit}
Z[A;y,h]= \frac{1}{J[A]} \times Z_{\text{light}}[A] \times Z_{\text{mirror}}[A;y,h]~.
\eeq  
We now make a couple of comments about (\ref{eq:Zsplit}) and refer the reader to \cite{Poppitz:2007tu,Poppitz:2009gt} for details of the derivation and further discussion:
\begin{enumerate}
\item The light partition function $Z_{\text{light}}[A]$ can be written in a closed form, as the corresponding integral over light fermions is Gaussian:\footnote{We hope the notation is not too confusing: $D[q_C A_\mu]$ denotes the GW operator as a functional of $q_C A_\mu$, where $q_C$ is the charge of the $C$ fermion and $A_\mu$---the gauge background; a similar statement holds for the $\hat\gamma_5$ eigenvectors and their $q_\alpha A_\mu$ dependence.}
\begin{eqnarray}
\label{eq:Zlight_total}
Z_{\text{light}}[A]&& \\
&=& \det \parallel w^\dagger_{iC}[q_C A_\mu] \cdot D[q_C A_\mu] \cdot t_{j} \parallel
 \prod_{\alpha=A,B} \det \parallel u^\dagger_{i\alpha}[q_\alpha A_\mu]
\cdot D[q_\alpha A_\mu]\cdot v_{j} \parallel ,\nonumber
\end{eqnarray}
where in each case the determinant is over the indices $i,j$ labeling the eigenvectors and 
we omitted the contribution of $X_-$ to $Z_{\text{light}}$ as it does not introduce an $A_\mu$ dependence.
The  $q_\alpha$ corresponding to $A$, $B$, $C$ are 3, 4, 5 respectively.  
\item The expression for the mirror partition function $Z_{\text{mirror}}[A;y,h]$ cannot
be given analytically in a compact form such as \myref{eq:Zlight_total}. Because  
of the Yukawa interactions, the mirror partition function is defined by a highly nontrivial integral over mirror fermions and  $\phi$.
\item The Jacobian $J[A]$ arises from the change of variables from the $\psi, \bar\psi$-basis in (\ref{v11}) to the eigenvector basis (\ref{eq:hatgamma5},\ref{eq:gamma5}), 
and is a product of Jacobians   for each of the fields $A$, $B$ and $C$; again
we ignore an $A_\mu$-independent Jacobian for the neutral spectator $X$. The explicit form of the Jacobian is  given in \cite{Poppitz:2007tu, Poppitz:2009gt}. The variation of the Jacobian $J[A]$ with respect to the gauge field  depends solely on the variation of the $\hat\gamma_5$ eigenvectors, which 
introduces the so-called  ``measure terms'' (whose role in defining purely chiral path integrals is crucial and  constitutes the unsolved ``measure problem"). However,  in our vector-like theory (\ref{v11}) these can be ignored as the ``measure terms" of the light and mirror sectors exactly cancel  and contain no dynamical information, see \cite{Poppitz:2009gt} for a detailed discussion. 
\end{enumerate}

\subsection{Properties of the mirror polarization tensor}

\label{polarization}
\subsubsection{Definition}
To see whether the mirror degrees of freedom are decoupled (i.e., have mass of order the inverse lattice spacing) 
or not, we will  study the analytic behavior of 
the photon vacuum polarization tensor.  It is defined as:
\beq \label{eq:Pi}
\left.\Pi_{\mu\nu}(x,y) \equiv \frac{\delta^2 \ln Z[A]}{\delta A(x) \delta A(y)} \right|_{A=0}.
\eeq
Just as the partition function  splits, recall  (\ref{eq:Zsplit}), into the product of ``mirror" and ``light" parts, so does the polarization operator:
\beq 
\Pi_{\mu\nu}(x,y)=\Pi^{\text{light}}_{\mu\nu}(x,y) + \Pi^{\text{mirror}}_{\mu\nu}(x,y)~,
\label{eq:Pisplit}
\eeq
where the 
 mirror polarization tensor is explicitly defined  by:
\beq
\label{mirrorp1}
\Pi^{\text{mirror}}_{\mu\nu} \equiv \left. \frac{\delta^2 \ln Z_{\text{mirror}}[A]}{\delta A(x) \delta A(y)} \right|_{A=0}~.
\eeq
The contribution of the Jacobian, following the same
convention as in \cite{Poppitz:2009gt}, is included into $\Pi^{light}_{\mu\nu}(x,y)$.

\subsubsection{The splitting theorem}
The ``splitting theorem'' developed in \cite{Poppitz:2007tu} makes the anomaly matching 
idea clear on lattice and at the same time serves as an indispensable tool to 
calculate the polarization tensor.
It states that under an arbitrary variation of the background gauge field, 
$A_\mu \to A_\mu + \delta A_\mu$, 
the change of the 
chiral partition function $Z_{\text{mirror}}$ is the sum of two parts:
\beq \label{eq:split_th}
\delta \ln Z_{\text{mirror}}[A]
=\sum_i \[ (\delta w^\dagger_{iA} \cdot w_{iA}) + (\delta w^\dagger_{iB} \cdot w_{iB} ) 
+ (\delta u_{iC}^\dagger \cdot u_{iC}) \] + \left\langle \frac{\delta S}{\delta O} \delta O \right\rangle~,
\eeq
where the brackets in the last term denote an expectation value calculated 
with the partition function $Z_{\text{mirror}}[A]$. Also in the last term in (\ref{eq:split_th}), 
$O$  denote the operators ($D[qA]$ and  
$\hat P_{q,\pm}$) entering the mirror action and $\delta{O}$---their variation with the gauge background; a sum over all operators and their variations is implied.
The first three terms  in (\ref{eq:split_th}), instead, depend solely on the variation of the $\hat\gamma_5$ eigenvectors with the gauge background and 
 define the  ``measure currents'': 
\beq
 \sum\limits_i (\delta w^\dagger_{iA} \cdot w_{iA}) \equiv j^{w_A}_\mu \delta A_\mu,  \quad
\sum\limits_i (\delta u^\dagger_{iC} \cdot u_{iC}) \equiv j^{u_C}_\mu \delta A_\mu~,\nonumber
\eeq
as well as an expression for $j_\mu^{w_B}$ identical  to that for $j_\mu^{w_A}$ given above with $A \rightarrow B$.

Since the splitting theorem (\ref{eq:split_th}) gives the variation of any chiral partition function, its repeated application is useful in the 
  calculation of the explicit expression of $\Pi_{\mu\nu}$ (\ref{mirrorp1}) of the mirror theory for the 3-4-5 model. 
The splitting theorem implies that one can further decompose the mirror polarization tensor into a part that arises from the variation of the measure currents and a remainder part, 
denoted by $\Pi^{\text{mirror},\prime}_{\mu\nu}$.  With the abbreviation (to be used extensively below) $\delta_\mu\equiv\delta/\delta A_\mu$, where the subscript $\mu$ includes both the gauge-field vector index and the space-time position,  Eqs.~(\ref{eq:split_th}) and (\ref{mirrorp1}) imply that \cite{Poppitz:2009gt}:
\beq 
\label{eq:Pimirrorsplit}
\Pi^{\text{mirror}}_{\mu\nu}= \delta_\nu j^{w^A}_\mu
+ \delta_\nu j^{w^B}_\mu + \delta_\nu j^{uC}_\mu + \Pi^{\text{mirror},\prime}_{\mu\nu}. 
\eeq
We stress that $\Pi^{\text{mirror},\prime}_{\mu\nu}$ comes from the variation of the interaction terms in the mirror sector and therefore ``knows" about the mirror dynamics, while 
the first three measure current terms are lattice artifacts---they come from the variation of 
the eigenvectors of the lattice operator $\hat\gamma_5$, are independent of the dynamics (furthermore, their contribution is cancelled by the corresponding terms in the light polarization operator).  In our numerical work,
we have calculated $\Pi^{\text{mirror}\prime}_{\mu\nu}$ as it represents  the
part of (\ref{eq:Pimirrorsplit}) that contains the information on mirror dynamics.

\subsubsection{Transversality and symmetry}

\label{transversality}

There are two exact identities obeyed by $\Pi^{\text{mirror},\prime}_{\mu\nu}$ which
we now review.  These identities hold for arbitrary anomaly-free mirror partition functions,  are  independent of the  interaction strength, and  
thus serve as an important check of our measurement code.
 
We begin by noting that the total partition function $Z[A]$ is gauge invariant,
$Z[A+\delta_\omega A]=Z[A]$, where $\delta_\omega A_\mu=- \nabla_\mu \omega$ and 
$\nabla_\mu \omega=\omega_{x+\mu}-\omega_x$. This implies the
transversality of all $n$-point functions,
\beq \label{eq:trans-npoint}
\sum_\mu \left. \nabla^*_{\mu_1 x_1} \frac{\delta^n \ln Z[A]}
{\delta A_{\mu_1}(x_1) \cdots \delta A_{\mu_{n}}(x_{n})} \right|_{A_\mu = 0}~.
\eeq
As a particular case, the polarization tensor (\ref{eq:Pi}) of the full (light $+$ mirror) theory satisfies the transversality 
condition:
\beq 
\sum_\mu \nabla^{*}_{\mu x} \Pi_{\mu\nu}(x,y) = 0.
\label{eq:transPi} 
\eeq
There  is no a priori reason to assume that on the lattice 
this condition is satisfied separately by $\Pi^{\text{light}}_{\mu\nu}$ 
and $\Pi^{\text{mirror}}_{\mu\nu}$. However, in the case of an anomaly free mirror theory, the following exact---i.e., independent on the mirror couplings---properties of the mirror polarization tensor $\Pi_{\mu\nu}^{\text{mirror} \prime}$ (\ref{eq:Pimirrorsplit}) follow: 
\begin{enumerate}
\item Because the light and mirror theories are separately anomaly free, the transversality of the total polarization operator implies that also:
\beq 
\label{trans_mirror}
\sum_{\mu} \nabla^*_{\mu x} \Pi^{\text{mirror},\prime}_{\mu\nu}(x,y) = 0, 
\eeq
i.e., the mirror polarization operator $\Pi_{\mu\nu}^{\text{mirror} \prime}$ is transverse w.r.t. the first index (we have momentarily restored explicit position dependence). Eq.~(\ref{trans_mirror}) follows from anomaly freedom and various identities obeyed by the  measure current contributions to the polarization operator and the anomaly, see \cite{Poppitz:2009gt}.

\item Furthermore, the antisymmetric part of $\Pi_{\mu\nu}^{\text{mirror} \prime}$ obeys   \cite{Poppitz:2009gt}:
\beq
\label{pi1}
\Pi^{\text{mirror},\prime}_{\mu\nu} - \Pi^{\text{mirror},\prime}_{\nu\mu}
&=& - (\delta_\nu j^{u_C}_\mu - \delta_\mu j^{u_C}_\nu) 
 - \sum_{\alpha=A,B} ( \delta_\nu j^{w_\alpha}_\mu - \delta_\mu j^{w_\alpha}_\nu  ) \nonumber \\
 &=& \sum_{\alpha=A,B,C} \mathcal F^\alpha_{\mu\nu} .
\eeq
The quantity $\mathcal F_{\mu\nu}^\alpha$ is the curvature of the measure current whose 
divergence $\nabla^*_\mu \mathcal F_{\mu\nu}^\alpha$ gives ``half" the anomaly.   $\mathcal F_{\mu\nu}^\alpha$ is a known functional of the gauge background: 
\beq
\label{fexplicit}
\mathcal F_{\mu\nu}^\alpha  = - \Tr \hat{P}_\alpha \[ \delta_\mu\hat{P}_\alpha, \delta_\nu\hat{P}_\alpha \]~,
\eeq
where $\hat{P}_{\alpha = A,B}[A_\mu] = \hat{P}_-[q_{\alpha = A,B} A_\mu]$ and $\hat{P}_{\alpha = C}[A_\mu] = \hat{P}_+[q_C A_\mu]$. Furthermore, at vanishing gauge background (for brevity, we do not indicate this explicitly below, as well as the fact that each $\mathcal F_{\mu\nu}^\alpha$ is a functional of $q_\alpha A_\lambda$, which is important to obtain (\ref{fexplicit1})) we have, recalling that $\hat{P}_+ + \hat{P}_- = 1$:
\begin{eqnarray}
\label{fexplicit1}
  \sum\limits_{\alpha=A,B,C} \mathcal F^\alpha_{\mu\nu}&=& - (3^2 + 4^2) \Tr \hat{P}_- \[ \delta_\mu\hat{P}_-, \delta_\nu\hat{P}_- \] - 5^2 \;
 \Tr \hat{P}_+ \[ \delta_\mu\hat{P}_+, \delta_\nu\hat{P}_+ \] \nonumber \\ 
 &=& - 5^2 \Tr  (\hat{P}_- + \hat{P}_+) \[ \delta_\mu\hat{P}_-, \delta_\nu\hat{P}_- \]  =0.
\end{eqnarray}
Thus, in the 3-4-5 model, the sum of the
different $\mathcal F^\alpha_{\mu\nu}\vert_{A=0}$ vanishes
  since it is an anomaly free theory. 
Together with (\ref{pi1}), this  implies that  $\Pi_{\mu\nu}^{\text{mirror} \prime}[0]$ is symmetric upon interchange of $\mu$ and $\nu$ (which includes interchange of spacetime coordinates). We note that symmetry also implies that $\Pi_{\mu\nu}^{\text{mirror} \prime}$ is transverse also w.r.t. the second index.\footnote{Transversality  w.r.t. second index is  a property satisfied by $\Pi'_{\mu\nu}$ for any chiral partition function, not necessarily anomaly-free, see Appendix A of \cite{Poppitz:2009gt}. However, in the anomaly-free case it follows from transversality w.r.t. first index and symmetry.}

\end{enumerate}
 
We have used transversality (\ref{trans_mirror}) and symmetry (\ref{fexplicit}, \ref{fexplicit1})  
as  consistency checks
on our computer code for the calculation of $\Pi^{\text{mirror},\prime}_{\mu\nu}$,
which is about 3,000 lines long.  Results of these checks will
be presented in Section \ref{tests}. 

\subsection{Derivation of mirror polarization tensor}
\label{deriv}
The set up for calculating $\Pi^{\text{mirror},\prime}_{\mu\nu}$ is a straightforward
generalization of what was done for the 1-0 model in 
\cite{Poppitz:2009gt}.  We first expand the mirror fermion fields onto the chiral eigenvectors:
\beq 
&& \bar A_- = \sum_i \bar\alpha_-^i w_{iA}^\dagger, \quad
\bar B_- = \sum_i \bar\beta_-^i w_{iB}^\dagger, \quad
\bar C_+ = \sum_i \bar\gamma_+^i u_{iC}^\dagger, \quad
\bar X_+ = \sum_i \bar\chi_+^i u_{iX}^\dagger, \nnn
&& A_- = \sum_i \alpha_-^i t_i, \quad B_- = \sum_i \beta_-^i t_i, \quad
C_+ = \sum_i \gamma_+^i v_i, \quad X_+ = \sum_i \chi_+^i v_i,
\label{eqme}
\eeq
where $\alpha^i_-, \beta^i_-, \gamma^i_+, \chi^i_+$ and their conjugates are the Grassmann variables of integration in the path integral used to  define the mirror partition function. Upon the change of variables (\ref{eqme}), 
the fermion kinetic part of $S_{\text{mirror}}$, see \myref{eq:action}, takes the form:
\beq
S_{\text{mirror,kin.}} &=& -\bar \alpha_{-}^i \alpha_{-}^j (w_{iA}^\dagger \hat P_{+A} D_3 t_{j})
-\bar \beta_{-}^i \beta_{-}^j (w_{iB}^\dagger \hat P_{+B} D_4 t_{j})
\ddd -\bar \gamma_{+}^i \gamma_{+}^j (u_{iC}^\dagger \hat P_{-C} D_5 v_{j}).
\eeq
The interaction terms \myref{eq:action} include the Dirac part:
\beq \label{eq:yuk.dirac}
S_{\text{Yuk.},\text{Dirac}}
&=& y_{30} \bar\alpha_-^i \chi_+^j (w_{iA}^\dagger \cdot \hat P_{+A} \cdot \phi^{-3} \cdot v_j)
+ y_{40} \bar\beta_-^i \chi_+^j (w_{iB}^\dagger \cdot \hat P_{+B} \cdot \phi^{-4} \cdot v_j)
\ddd + y_{35} \bar\alpha_-^i \gamma_+^j (w_{iA}^\dagger \cdot \hat P_{+A} \cdot \phi^{2} \cdot v_j)
+ y_{45} \bar\beta_-^i \gamma_+^j (w_{iB}^\dagger \cdot \hat P_{+B} \cdot \phi \cdot v_j)
\ddd + y_{30} \bar\chi_+^i \alpha_-^j (u_{iX}^\dagger \cdot \hat P_{-X} \cdot \phi^3 \cdot t_j)
+ y_{40} \bar\chi_+^i \beta_-^j (u_{iX}^\dagger \cdot \hat P_{-X} \cdot \phi^4 \cdot t_j)
\ddd + y_{35} \bar\gamma_+^i \alpha_-^j (u_{iC}^\dagger \cdot \hat P_{-C} \cdot \phi^{-2} \cdot t_j)
\ddd + y_{45} \bar\gamma_+^i \beta_-^j (u_{iC}^\dagger \cdot \hat P_{-C} \cdot \phi^{-1} \cdot t_j),
\eeq
and a Majorana part:
\beq \label{eq:yuk.maj}
S_{\text{Yuk.},\text{Maj.}}
&=& -h_{30} \bar\chi_+^i \bar\alpha_-^j (u_{iX}^\dagger \cdot \phi^{-3}
\gamma_2 \cdot \hat P_{+A}^T \cdot w_{jA}^* )
-h_{40} \bar\chi_+^i \bar\beta_-^j (u_{iX}^\dagger \cdot \phi^{-4}
\gamma_2 \cdot \hat P_{+B}^T \cdot w_{jB}^* )
\ddd -h_{35} \bar\gamma_+^i \bar\alpha_-^j (u_{iC}^\dagger \cdot \hat P_{-C} \cdot \phi^{-8}
\gamma_2 \cdot \hat P_{+A}^T \cdot w_{jA}^* )
\ddd -h_{45} \bar\gamma_+^i \bar\beta_-^j (u_{iC}^\dagger \cdot \hat P_{-C} \cdot \phi^{-9}
\gamma_2 \cdot \hat P_{+B}^T \cdot w_{jB}^* ) + U\text{-independent}.
\eeq
The dots in the equations above denote contractions of both spacetime position and spinor indices.
Lastly, we have the kinetic terms of the unitary Higgs field $S_\kappa$, 
given above in Eq.~\myref{Sk}.  In our study we have set
$\kappa=1/2$, since this corresponds to the symmetric phase. 
Setting kappa to a much larger value would lead to algebraic
ordering, and would break the chiral symmetry in the mirror sector 
spontaneously, resulting in the existence of a Green-Schwarz field \cite{Poppitz:2009gt}.  

Performing the functional derivatives in $\Pi_{\mu\nu}$ (\ref{eq:Pi}) and using the splitting theorem many times, as in \cite{Poppitz:2009gt}, we have
derived the explicit expression for $\Pi^{{\text{mirror}}}_{\mu\nu}$.
This lengthy expression is given in Appendix \ref{exppi}.
It can be seen that it does take the general form  (\ref{eq:Pimirrorsplit}) 
that was given above, with a rather long expression for $\Pi^{{\text{mirror}}'}_{\mu\nu}$.  The quantities such as $\vev{\bar\alpha_-^i \chi_+^j }$
will correspond to expectation values of the inverse of the fermion
matrix that is derived from the action above, after having
implemented the change of variables \myref{eqme}.  Explicit
expressions for the matrix are given in Appendix \ref{fmat}.
It can be seen from the expression in Appendix \ref{exppi}
that the $A_\mu$ dependence of the projection
operators leads to many terms with $\delta_\mu \hat P_\pm$, due
to the chiral Yukawa couplings.  This proliferation of gauge field
dependence in the lattice theory yields an expression that is a challenge to
correctly implement in computer code, hence the importance of the
consistency checks listed in Section \ref{transversality}.

\section{Numerical studies}
\label{numer}
In all of our studies we are working in the limit of very
large Yukawa couplings, where the kinetic terms can be
neglected.  This leads to a simplification of the fermion
matrix, so that in all of our discussion below the matrix $M$
is only the non-vanishing part in this limit.  Also,
many terms in the expression for the polarization tensor
given in Appendix \ref{exppi} vanish in this limit.
Essentially, there must be an equal number of $+$
signs as $-$ signs on the fermion fields entering
expectation values.  Thus, $\vev{ \bar \alpha_{-}^i \alpha_{-}^j }=0$
whereas $\vev{\bar\alpha_-^i \chi_+^j } \not= 0$.
All Yukawa couplings are subsequently rescaled by
an overall factor so that $y_{q_L q_R}=\ord{1}$ in
our calculations.  The behavior of the polarization
tensor is independent of this overall factor.

As mentioned above, in all of our studies we take the hopping parameter in the Higgs field
action to be $\kappa=1/2$.  We have verified that this is in the
symmetric phase, by measuring the scalar susceptibility
and checking that it is independent of the number of
lattice sites, $N \times N$.

\subsection{Monte Carlo simulation}
In order to have efficient sampling for the scalar field $\phi$,
we have applied the Wolff cluster updating algorithm \cite{Wolff:1988uh} to the
action \myref{Sk}.  Fermions are taken into account by explicitly
calculating the determinant of the fermion matrix $M$, and then
reweighting the scalar field configurations.  Thus, the partition
function is:
\beq
Z_{\text{mirror}} = \int [d\phi] \det M(\phi) e^{S_\kappa}~,
\eeq
and expectation values of operators $\Ocal(\phi)$ (typically
elements of $M^{-1}$) are computed according to:
\beq
\vev{ \Ocal(\phi) } = \frac{ \vev{ \Ocal(\phi) \det M(\phi) }_\phi }{ \vev{ \det M(\phi) }_\phi }~.
\eeq
Here, $\vev{ \cdots }_\phi$ indicates an average over the $\phi$ configurations
generated with the Wolff algorithm.  For example,
\beq
\vev{ \det M(\phi) }_\phi = \frac{1}{N} \sum_{i=1}^N \det M(\phi_i),
\eeq
where the configurations $\phi_i$ are taken from the canonical
distribution:
\beq
P(\phi) = \frac{ e^{S_\kappa(\phi)} }{ \int [d\phi'] e^{S_\kappa(\phi')} }~.
\eeq
We separate samples by 40 cluster updates, which we find to
give autocorrelation times much less than unity, so that our
samples are statistically independent.  We find that that
$\ord{1000}$ samples give stable averages for the quantities
we measure.  The computational time required for the simulation,
including the calculation of the fermion determinant, is negligible
compared to the time that is required for our measurement of
the polarization tensor.  That is because the latter requires
many nested loops of sums over momenta.

\subsection{Complex phase}
\label{cph}
\begin{figure}
\centering
\includegraphics[width=4in,height=3in]{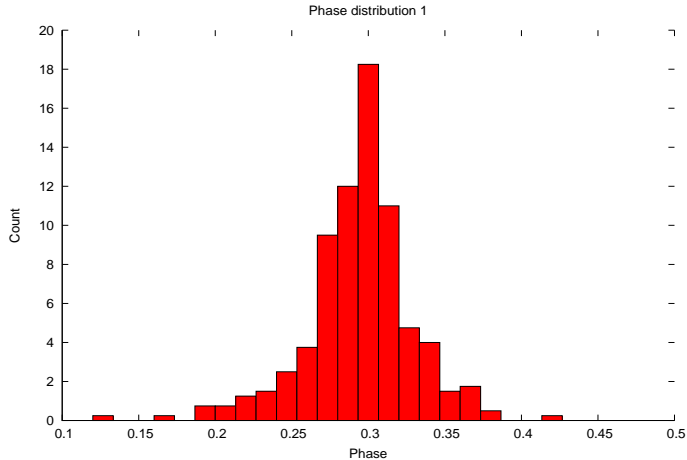} 
\caption{Complex phase distribution 
for the set of Yukawa coupling constants given
in Table \ref{tb:coupling1}, on an $8 \times 8$ lattice. \label{fig:phase1}}
\end{figure}

The fermion determinant of our theory is not real.  Thus there is a potential
problem of a complex phase leading to large cancellations and hence large
statistical errors, since the signal would be much smaller than fluctuations.
To overcome this feature, we have studied the dependence of the phase
distribution on the values of the Yukawa coupling parameters.  We have optimized
these parameters in order to have a narrow phase distribution, so that
cancellations are minimal.  An example of one of our phase distributions
is shown in Figure \ref{fig:phase1}, corresponding to the set of
parameters shown in Table \ref{tb:coupling1}.  It can be seen that it is
reasonably narrow.  Nevertheless since we calculate the fermion determinant
at each sampling step of our simulation, we include the complex phase
in our expectation values.

\begin{table}
\centering
\begin{tabular}{cccccccc}
\hline
$h_{30}$ & $h_{40}$& $h_{35}$& $h_{45}$& $y_{30}$&$y_{40}$ & $y_{35}$& $y_{45} $ \\
\hline
   30.3214 & 3.08123 & 3.00278 & 23.7109 & 1.0 & 1.0 & 1.0 & 1.0  \\
\hline
\end{tabular}
\caption{A set of Yukawa coupling constants that we found by
optimizing narrowness of the complex phase distribution for an $8 \times 8$ lattice.  
 \label{tb:coupling1}}
\end{table}

\subsection{Tests of the polarization tensor code}
\label{tests}
We tested the polarization tensor code with the transversality condition,
which in momentum space becomes:
\beq
\sum_\mu (1-\omega^{k_\mu}_N) \tilde \Pi_{\mu\nu}^{\text{mirror},\prime}(k) = 0~,
\label{trksp}
\eeq
where $N$ is the number of lattice sites in each direction
on the $N \times N$ lattice, $\omega_N \equiv e^{2 \pi i \over N}$ and 
the Fourier transform of $\Pi_{\mu\nu}^{\text{mirror},\prime}(x,y)$ is defined in the usual manner:
\beq
\label{fourierpi}
\tilde \Pi_{\mu\nu}^{\text{mirror},\prime}(k) 
= \sum\limits_{x_1,x_2=1}^N \omega_N^{k_1 x_1 + k_2 x_2}\; \Pi_{\mu\nu}^{\text{mirror},\prime}(x,0)~.
\eeq
Subsequently, we will abbreviate $\tilde\Pi_{\mu\nu}^{\text{mirror},\prime}(k) \to \Pi_{\mu\nu}(k)$.
In the case of $k_1=k_2$, transversality requires that the $\nu=1$ components 
are related by $\Pi_{21} = -\Pi_{11}$.
The results of this test on a $6 \times 6$
lattice are given in Table \ref{transcheck1},
where we used the set of couplings given in Table \ref{tb:coupling2} below.
First it can be
seen that for $k_1=k_2$ the imaginary part of the polarization tensor is zero to within the
statistical errors.  Secondly, $\Pi_{11}$ and $-\Pi_{21}$ are nearly
equal to each other in this case.  We present the result of Eq.~\myref{trksp} in the
last column, and it can be seen that it is within $\sim 1\sigma$ of zero.   
Thus the transversality is satisfied to within statistical errors.
We have also considered cases where $k_1 \not= k_2$.  Here again,
the results of Eq.~\myref{trksp} are within $\sim 1\sigma$ of zero,
consistent with transversality.

\begin{table}\begin{center}
\begin{tabular}{|c|c|c|c|} 
\hline\hline
$(k_1,k_2)$ & $\Pi_{11}$ & $\Pi_{21}$ & $\nabla_\mu \Pi_{\mu\nu}$ \\ 
\hline\hline
(1,1) & $-1.607(16) $ & $1.615(18) $
  & $6(14) \times 10^{-3}$ \\ 
  &$  + i \;0.0007(97)$ & $- i \; 0.0014(65)$ &  $- i \; 6(22)  \times 10^{-3}$\\ 
  \hline
(1,5) & $-1.602(17)  $ & $-0.819(10)  $ 
  & $ -1.3(1.8) \times 10^{-2}$ \\ 
    & $  - i \; 0.004(10)$ & $ - i  \;1.386(15)$ 
  & $ -i \; 1.7(2.0) \times 10^{-2}$ \\ \hline
(2,4) & $1.498(23)  $ & $-0.738(10)  $ 
  & $ 36(43)   \times 10^{-3}$ \\
  & $+  i \; 0.0032(40)$ & $+  i  \;1.279(16)$ 
  & $  - i  \;1.2(2.7) \times 10^{-2}$ \\ \hline
\hline
\end{tabular}
\caption{Transversality check for $6 \times 6$ lattice, with respect to the first index.  In
each case the sample size is 10,000 configurations.  Statistical errors are quoted
in parentheses.  Transversality it satisfied to within errors.  
\label{transcheck1} }\end{center}
\end{table}

Results where transversality in the second index is probed are shown
in Table \ref{transcheck2}.  Once again, from the last column one
can see that all values for $\sum_\nu (1-\omega^{-k_\nu}_N) \tilde \Pi_{\mu\nu}$
are within $\sim 1 \sigma$ of zero, statistically equivalent
to a vanishing result.

\begin{table}\begin{center}
\begin{tabular}{|c|c|c|c|}
\hline\hline
$(k_1,k_2)$ & $\Pi_{11}$ & $\Pi_{12}$ & $\nabla_\nu \Pi_{\mu\nu}$ \\ 
\hline\hline
(1,5) & $-1.602(17) $ & $-0.801(10)  $ 
  & $1.1(1.4) \times 10^{-2}$ \\
  & $  - i \; 0.004(10)$ & $+  i \; 1.396(15)$ 
  & $  +i \; 3(23)\times 10^{-3}$ \\ \hline
(2,4) & $1.490(22) $ & $-0.719(10)  $ 
  & $5.4(4.3)  \times 10^{-2}$ \\
 & $+i\; 0.0050(36) $ & $-i\; 1.266(15)  $ 
  & $+i \; 2.1(2.7)  \times 10^{-2}$ \\ \hline
(1,1) & $1.607(16) $ & $-1.615(17)  $ 
  & $5(14)  \times 10^{-3}$ \\
  & $  +i \;0.0007(97)$ & $+  i  \;0.0011(61)$
  & $  - i \; 6(22) \times 10^{-3}$ \\ \hline
\hline
\end{tabular}
\caption{Transversality check for $6 \times 6$ lattice, with respect to the second index.  In
each case the sample size is 10,000 configurations.  Statistical errors are quoted
in parentheses. Transversality it satisfied to within errors.  
\label{transcheck2} }\end{center}
\end{table}

Finally, in Table \ref{antitab} we show a check that the antisymmetric
part of the polarization tensor vanishes.  The result in the last
column is supposed to vanish within statistical error.  Four out of
six results deviate from zero by less than $1 \sigma$.  Two out of
six results (33\%) deviate by about $1.5\sigma$.  Statistically,
deviations by $1.5\sigma$ or more should occur about 15\% of the time.
We regard the results as being consistent with zero, though not perfectly
so.

\begin{table}\begin{center}\begin{tabular}{|c|c|c|c|}
\hline\hline
$(k_1,k_2)$ & $\Pi(k)_{21}$ & $\Pi(-k)_{12}$ & $\Pi(k)_{21}-\Pi(-k)_{12}$ \\ 
\hline\hline
(1,5) & $-0.819(10) $ & $-0.814(10) $
  & $-5(14) \times 10^{-3}$ \\
  & $ - i \; 1.386(15)$ & $ - i \; 1.421(15)$
  & $  +i  \;3.4(2.2) \times 10^{-2}$ \\ \hline
(2,4) & $-0.738(10)  $ & $-0.728(10)  $ 
  & $ -10(14)  \times 10^{-3}$ \\
 & $ + i  \;1.279(16)$ & $ + i  \;1.244(16)$ 
  & $ + i  \;3.5(2.3) \times 10^{-2}$ \\ \hline
(2,5) & $-0.0040(68)  $ & $0.0061(72)  $
  & $ -2.2(9.7) \times 10^{-3}$ \\ 
   & $ - i  \;0.139(16)$ & $ - i  \;0.128(16)$
  & $  - i  \;1.1(2.3) \times 10^{-2}$ \\ \hline
\hline
\end{tabular}
\caption{Antisymmetric part of the polarization tensor, for the $6 \times 6$ lattice.
The results are consistent with zero. \label{antitab} }\end{center}
\end{table}

We have also conducted these tests for the set of couplings given in Table \ref{tb:coupling1},
finding similar results.
The conclusion of these checks is that the simulation and measurement code
appears by all measures to be correct.  We have found that if just one term
in the expression for the polarization tensor is incorrect then these
checks are wildly different from zero.  We are seeing transversality
and vanishing antisymmetric part that is satisfied at the per cent level
relative to the size of $\Pi_{\mu\nu}$.  On this basis we are confident
in the results for $\Pi_{\mu\nu}$ which we will present next.

\subsection{Probing for massless particles}
In the continuum, the contribution to the Fourier transform 
of the real part of the polarization tensors due to 
massless particles takes the form:
\beq 
\label{eq:RePi}
\tilde\Pi_{\mu\nu}(k)= 2C \frac{\delta_{\mu\nu}k^2-k_\mu k_\nu}{k^2}~.
\eeq 
Two specific examples of contributions to $C$ were discussed in \cite{Poppitz:2009gt} (the results (\ref{eq:Cgs},\ref{eq:Cgw}) are derived there). For a free Green-Schwarz scalar field, 
which shifts under the gauge symmetry\footnote{This field could also be called a ``St\" uckelberg field", but the term ``Green-Schwarz" was used in the 1-0 model, since it reproduces the mirror sector anomaly via its Wess-Zumino coupling to the photon.}:
\beq \label{eq:Cgs}
2C_{\text{GS scalar}} \simeq -\kappa q^2~,
\eeq
where $\kappa$ is the coefficient of its kinetic term (similar to our $S_\kappa$, see (\ref{Sk})).
On the other hand, a single free charge-$q$ Weyl fermion, gives  a contribution equal to that of a half charge-$q$ Dirac fermion:
\beq \label{eq:Cgw}
2C_{\text{fermion}} \simeq -\frac{1}{2\pi} q^2~.
\eeq
The way to look for a  massless pole in (\ref{eq:RePi}) is to  notice that (\ref{eq:RePi}) 
has a directional limit as $k \rightarrow 0$:
\beq \label{eq:direclimit}
\left.\tilde\Pi_{11}(\phi) \right|_{k\rightarrow 0}&=&C(1-\cos 2\phi),  \nonumber \\
\left.\tilde\Pi_{21}(\phi) \right|_{k\rightarrow 0}&=&-C \sin 2\phi,
\eeq
where $\phi$ is the angle of approach to origin measured from the positive-$k_1$ axis. 
Therefore if there is a massless particle in the spectrum
of the mirror theory, we would expect $\tilde\Pi_{\mu\nu}^{\text{mirror},\prime}$ 
has the following behavior as $k\rightarrow 0$:
\beq \label{eq:direclimit2}
\tilde\Pi_{11}(45^o)&=&-\tilde\Pi_{21}(45^o)=C~,   \nonumber \\
\tilde\Pi_{11}(90^o)&=&2C~, \nonumber \\
\tilde\Pi_{11}(0^o)&=&\tilde\Pi_{21}(0^o)=\tilde\Pi_{21}(90^o)~.   
\eeq

On the other hand, if there is no massless particle, we would expect:
\beq 
\tilde\Pi_{\mu\nu} \sim \frac{\delta_{\mu\nu} k^2-k_\mu k_\nu}{m^2}~,
\label{massive}
\eeq
as $k\rightarrow 0$.  Therefore if the mirror sector acquires
the desired mass scale, the directional 
limit behavior (\ref{eq:direclimit}) disappears. 

In the anomalous 1-0 mirror theory, Ref.~\cite{Poppitz:2009gt} found that the mirror polarization operator had precisely the discontinuity (\ref{eq:direclimit2}) with $C$ from (\ref{eq:Cgw}). The discontinuity obtained on an $8\times 8$ lattice was close to the continuum value of $C$ appropriate for a single unit-charge Weyl fermion---exactly as expected from the simplest solution of  't Hooft anomaly matching for unbroken gauge symmetry in the 1-0 model. As a check on the simulation, the 1-0 mirror theory was also driven into the ``broken" (algebraically ordered) phase by taking large $\kappa$ in (\ref{Sk}). Then, instead, a discontinuity (\ref{eq:Cgs}) appropriate to the Green-Schwarz field was observed, also remarkably close to the continuum value in the same $8 \times 8$ lattice. In the next Section, we  present the results of a similar study, this time for the anomaly-free 3-4-5 model.

\subsection{Results}
\subsubsection{Lattice size dependence of discontinuity of mirror polarization operator}
\label{latsize}
For the set of couplings given in Table \ref{tb:coupling1}, we
computed $\tilde\Pi_{\mu\nu}^{\text{mirror},\prime}(k)$ for
$6 \times 6$, $8 \times 8$ and $10 \times 10$ lattices.  The
$6\times 6$ lattice calculations could be performed on a 
desktop computer.  By contrast, the $8 \times 8$ and $10 \times 10$ 
lattices required the use of computing clusters. 
The reason for using such small lattices is because the calculation of 
$\tilde\Pi_{\mu\nu}^{\text{mirror},\prime}(k)$ 
is rather demanding and scales badly ($\sim N^{10}$ after
having taken into account momentum conservation) with
the number of lattice sites $N$ in each direction.
In particular we used over 200,000 core-hours
of computing in order to obtain our results.
We studied three directions in momentum space:
$0^o$ with ${\bf k}=(k,0)$; $45^o$ with ${\bf k}=(k,k)$;
$90^o$ with ${\bf k}=(0,k)$.  Here $k$ was taken
from $0,1,\ldots,N-1$ on an $N \times N$ lattice.
The corresponding momenta in physical units are $p=2 \pi k/Na$.

\begin{figure}
\centering
\includegraphics[width=4in,height=3in]{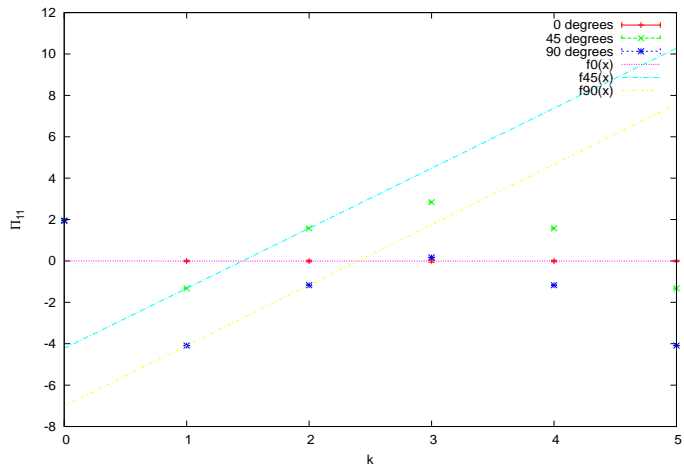}
\caption{$\tilde\Pi_{11}^{\text{mirror},\prime}(k)$ 
on a $6\times 6$ lattice.  The lines
show the extrapolation $k \to 0$ for different
angles of approach.  A clear discontinuity in
the directional limit can be seen.  \label{fig:Ne6}}
\end{figure}

Figure \ref{fig:Ne6} shows the results obtained on the $6 \times 6$ lattice.
It has the pattern suggested by (\ref{eq:direclimit}) with $C \approx 4 \approx 50/(4\pi)$.
In fact, it is remarkably similar to Figs.~2-5 of Ref.~\cite{Poppitz:2009gt},
except that the discontinuity is about 50 times larger, consistent
with contributions from the three charges that are present in the
underlying theory:  $3^2+4^2+5^2=50$.
The $k \to 0$ extrapolations are given in the second column of 
Table \ref{tb:fits}, where
we used a linear fit to the two smallest nonzero $k$ values.

\begin{table}
\centering
\begin{tabular}{cccc}
\hline
 $\phi$ & $A (6 \times 6)$ & $A (8 \times 8)$ & $A (10 \times 10)$ \\
\hline
 $0^o$ & -$2.36 \times 10^{-3}$ & -0.16423(82) & --- (0)\\
 $45^o$ & -4.22 & -3.820(98) & -3.764(41) \\
 $90^o$ & -7.01 & -6.43(16) & -6.501(47) \\
\hline
\end{tabular}
\caption{Linear extrapolation $k \to 0$, using a fit to $f(k) = A + B k$.
For the $6 \times 6$ lattice we use the two smallest nonzero values of $k$,
and so there is no error in the fit.  For the $8 \times 8$ and
$10 \times 10$ lattice we use the three smallest nonzero $k$ values,
and error in the fit is given in parentheses.  It can be seen
that the discontinuity is approximately constant in $N$. 
The $0^o$ value for $N=10$ was not calculated because of computer time constraints;  
based on the results from $N=6,8$, as well as calculations
in the free theory, we expect the $0^o$ values to vanish, within errors.
\label{tb:fits}}
\end{table}

\begin{figure} 
\centering
\includegraphics[width=4in,height=3in]{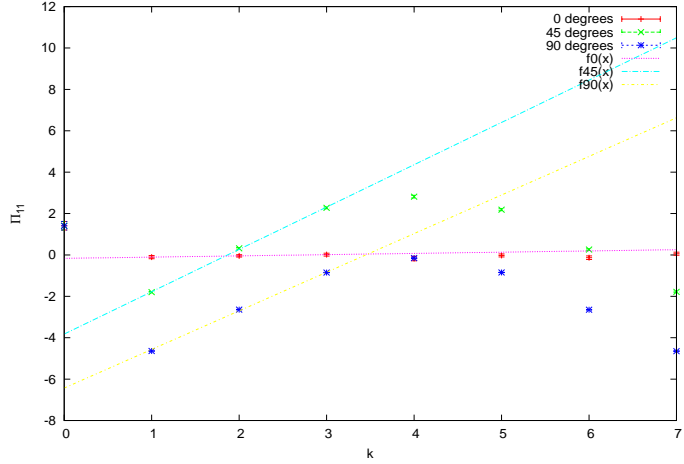}
\caption{$\tilde\Pi_{11}^{\text{mirror},\prime}$ on an $8 \times 8$ lattice
with the couplings given in Table \ref{tb:coupling1}. \label{fig:Ne8}}
\end{figure}

\begin{figure} 
\centering
\includegraphics[width=4in,height=3in]{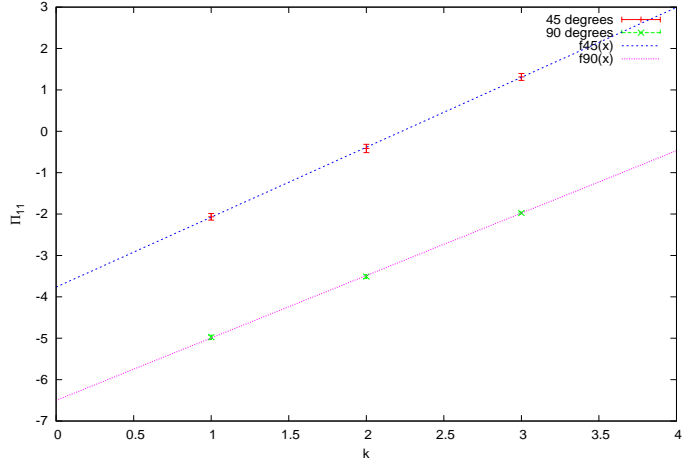}
\caption{$\tilde\Pi_{11}^{\text{mirror},\prime}$ on a
$10 \times 10$ lattice with the couplings given in Table \ref{tb:coupling1}.
Only the smallest values of $k$ were studied and the $0^o$ 
approach to the origin was omitted because of the expense
of the calculation.
\label{fig:Ne10}}
\end{figure}

Our results for an $8\times 8$ lattice with
the same Yukawa couplings are shown in Fig.~\ref{fig:Ne8}, with
$k \to 0$ extrapolations given in the third
column of Table \ref{tb:fits}.  As one can see, the discontinuity 
constant $C$ is approximately the same as was found on the $6 \times 6$ lattice.
Finally, on a $10 \times 10$ lattice we obtain the results
shown in Fig.~\ref{fig:Ne10}, with extrapolations provided in 
the final column of Table \ref{tb:fits}.
The discontinuity is again about the same.
 
\begin{figure} 
\centering
\includegraphics[width=4in,height=3in]{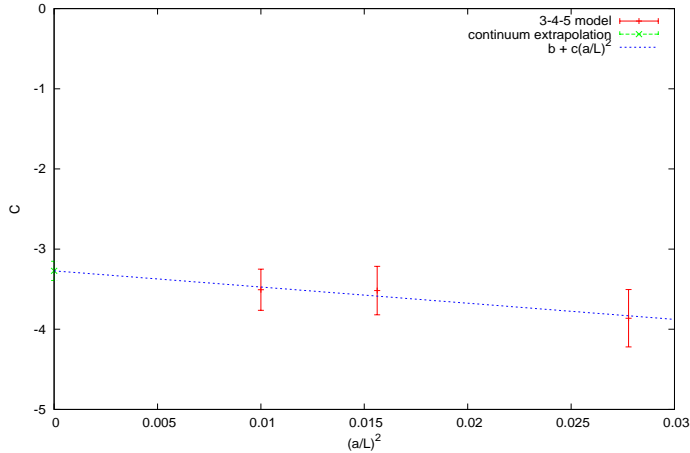}
\caption{The small-$k$ discontinuity in $\tilde{\Pi}_{11}$ for 
the 3-4-5 model mirror theory fitted to a linear function of $(a/L)^2$,
consistent with the known $\ord{a^2}$ errors of overlap fermions. 
The data from Table \ref{tb:fits} were used. 
The individual error bars are taken to be equal to 
half the difference of $C$ determined from the $45^o$ and $90^o$ 
fits.  It is seen that the discontinuity appears to approach a nonzero value as $a \to 0$. 
It is interesting to compare with a similar scaling performed with the same 
size lattices for a free massless Ginsparg-Wilson fermion in 
Figure \ref{fig:freemasslessdisc} and to a massive Ginsparg-Wilson 
fermion in Figure \ref{fig:massdiscfit}; more lattice sizes were 
used in the latter case. \label{fig:345discfit}}
\end{figure}

If the physical volume $L \times L$ is held fixed, where $L=Na$, while the number
of lattice sites $N \times N$ is increased, this corresponds to decreasing
the lattice spacing.  Thus the interpretation of the above results is that
the discontinuity is approximately constant as the lattice spacing is taken to smaller
values.  The discretization error of the overlap fermions which we are
using is known to be $\ord{a^2}$.  Thus the correct way to extrapolate
to the continuum limit is to fit the discontinuity to the functional
form:
\beq
C = b + c (a/L)^2 + \ord{(a/L)^4} = b + c N^{-2} + \ord{N^{-4}}~,
\eeq
and obtain the discontinuity $b$ in the continuum limit.  
Here, we use the values of $C$ from Table~\ref{tb:fits} for each of
the three values of $N=L/a$ that we have computed.  The errors
in the value of $C$ are dominated by the difference between the
value obtained from $45^o$ versus $90^o$.  Taking this into account,
we obtain the fit shown in Figure~\ref{fig:345discfit}.  It can
be seen that the linear function in $(a/L)^2$ describes the
data very well.  Furthermore, the linear extrapolation
intersects the $a=0$ limit at a nonzero discontinuity, $b= -3.27(12)$.  
This suggests that in the continuum
limit of the 3-4-5 model, the polarization tensor has a directional
discontinuity, consistent with having massless modes.  As remarked
above, the size of the discontinuity is about 50 times larger
than the 1-0 model, consistent with the charges of the fermions
in the underlying theory.

\begin{figure} 
\centering
\includegraphics[width=4in,height=3in]{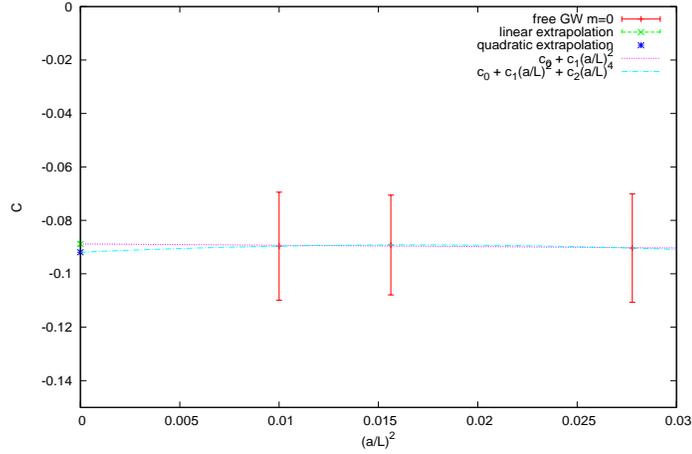}
\caption{The small-$k$ discontinuity of $\tilde{\Pi}_{11}$ for a 
free massless charge-one Weyl (one-half Dirac) Ginsparg-Wilson fermion
 as a function of $(a/L)^2$ for $N=6,8,10$, performed in a manner 
identical to that in Figure \ref{fig:345discfit}. 
The continuum value of the discontinuity is $\simeq 0.0796$, 
see (\ref{eq:Cgw}). \label{fig:freemasslessdisc}}
\end{figure}

\begin{figure} 
\centering
\includegraphics[width=4in,height=3in]{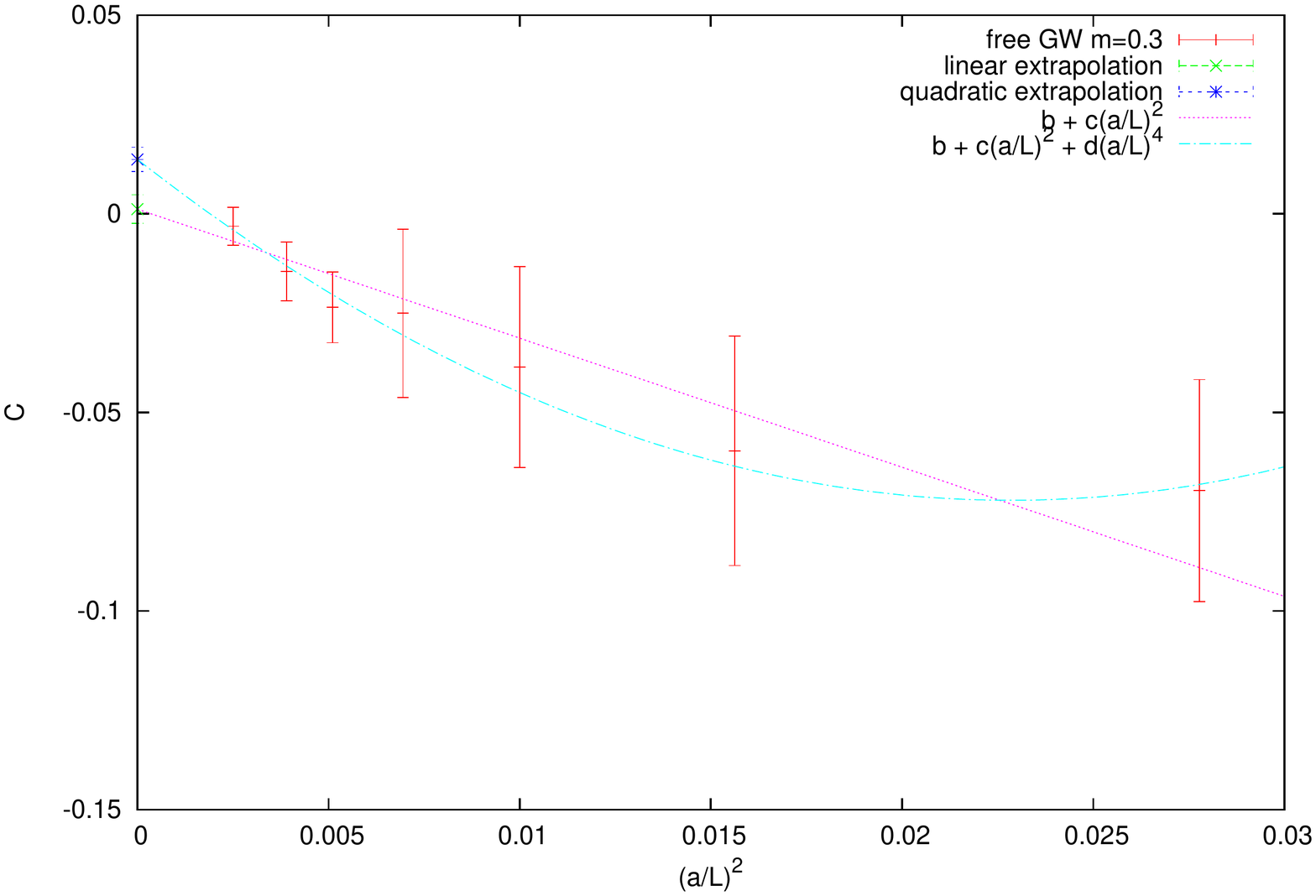}
\caption{The small-$k$ discontinuity in $\tilde{\Pi}_{11}$ for a 
massive unit-charge Weyl (one-half Dirac) fermion, of mass $m=0.3$ (in units of the lattice spacing).  
This value for $m$ was chosen to roughly match the discontinuity observed in the 3-4-5 model 
at the smallest lattices (divided by $50= 2 \times 5^2 =3^2 +4^2 + 5^2$).
It is seen that in the large-$N$ limit, 
there is no discontinuity, within the uncertainties of the extrapolation. \label{fig:massdiscfit}}
\end{figure}

The same scaling, but for a free massless unit-charge Weyl (half-Dirac) 
Ginsparg-Wilson fermion is shown on Figure~\ref{fig:freemasslessdisc}. 
The plot clearly shows that the discontinuity of the continuum limit, 
see (\ref{eq:Cgw}), is approached already on the rather small lattices 
used and that the behavior of a massless (free) fermion is quite similar
to what we found in the 3-4-5 model.
By contrast, we also compare the result of the 3-4-5 model, Figure~\ref{fig:345discfit},
with the same fit but for a free massive (once again, one-half Dirac) Ginsparg-Wilson fermion, 
shown on Figure~\ref{fig:massdiscfit}.  While it is clear that there is no 
discontinuity for a massive fermion, the precise continuum extrapolation 
value one obtains mildly depends on the extrapolation procedure.  Indeed, 
this gives a measure of the systematic uncertainty in the estimate for the discontinuity in
the continuum limit.  The plots for the free (massless or massive) Ginsparg-Wilson 
fermion were generated using an exact calculation of the zero-field charge-one polarization operator for a Weyl-fermion of chirality denoted by $\pm$, on an $N\times N$ lattice, given (in $x$-space) by:\beq
\label{piprime5}
\Pi^{\prime \pm}_{\mu \nu} = {1 \over 2}\; \delta_\nu {\rm Tr} D^{-1} \delta_\mu D \pm {1\over 4} \;{\rm Tr} \gamma_5 \delta_\mu \delta_\nu D \pm {1\over 4}\; \delta_\nu {\rm Tr} \left[ \delta_\mu D, \gamma_5 \right] D^{-1}~,
\eeq 
where $D$ is the Ginsparg-Wilson operator (the calculation of the relevant traces is performed using the finite-$N$ expansions of $D$ in powers of $A_\mu$ from Appendix B of \cite{Poppitz:2009gt}). Error bars were obtained, as in Figure~\ref{fig:345discfit}, by the difference
between the $45^o$ and $90^o$ extractions of the discontinuity $C$.

The result of these comparisons is the following.  The lattice discretization error at finite $a$
gives rise to an apparent discontinuity $C \not= 0$, even in the free  massive case.  
However, as the continuum
limit $a \to 0$ is taken, in the massive case this discontinuity disappears.  
We do not see this behavior in the 3-4-5 model.  Instead, the discontinuity of the mirror polarization operator in the 
 3-4-5 model looks like the discontinuity for a collection of 
free massless Ginsparg-Wilson fermions. These results indicate that the 3-4-5 model has
massless mirror sector modes and that the polarization tensor has a directional
discontinuity that survives the continuum limit.

 The discontinuity we find is roughly $50$ times larger than that of a single unit-charge chiral Ginsparg-Wilson fermion given in (\ref{eq:Cgw}). Anomaly freedom of the mirror sector theory allows, then, for two interpretations of our result. The massless charged mirror-fermion spectrum is  either that of the original mirror theory, i.e., a chiral $3_-$, $4_-$, $5_+$ massless fermion representation, or a massless vectorlike $5_-$ and $5_+$ fermion (the $5_-$ could, e.g., be a $\phi* 4_-$ composite while the opposite could be the original $5_+$ mirror fermion). Distinguishing between the two possible spectra requires more ``experimental" or theoretical input, as we discuss in the concluding Section.
 
\subsubsection{Yukawa coupling dependence}
\begin{figure}
\centering
\includegraphics[width=4in,height=3in]{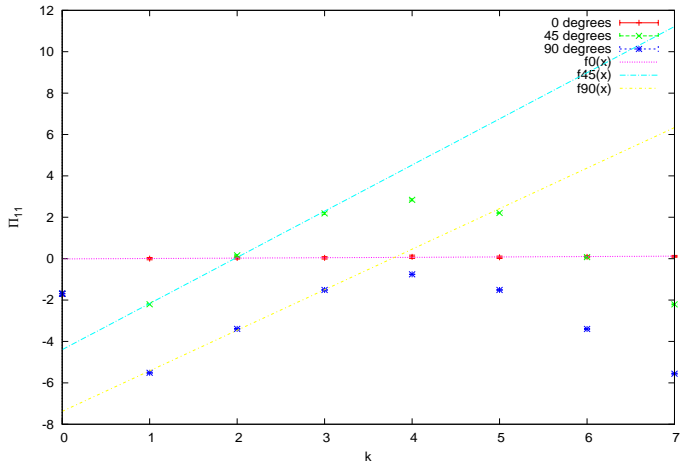}
\caption{The polarization tensor $\tilde\Pi_{11}^{\prime,\text{mirror}}$ 
on an $8 \times 8$ lattice with the set of couplings
given in Table \ref{tb:coupling2}.  A discontinuity approximately equal
to the one in Fig.~\ref{fig:Ne8} is seen, supporting the claim that the discontinuity
is largely independent of the Yukawa couplings.
\label{fig:Ne8_06072012}}
\end{figure}

The decoupling of the mirror fermions is an effect that is
supposed to occur for strong Yukawa couplings.  Thus it is
interesting to investigate a set of larger coupling constants.
In particular, we have studied how increasing the Majorana
Yukawa couplings, and taking $y \not = 1$, impacts the polarization tensor.  The 
new set of couplings
that we have studied are given in Table \ref{tb:coupling2}.
We have once again optimized the complex phase distribution,
as was discussed above, in order to have stable averages.
The results for an $8 \times 8$ lattice are shown in Fig.~\ref{fig:Ne8_06072012}.
Comparing to Fig.~\ref{fig:Ne8}, we see that to a good approximation
the discontinuity is independent of the Yukawa couplings.  Thus the
findings of the previous subsection appear to be quite robust.

\begin{table}
\centering
\begin{tabular}{c c c c c c c c}
\hline
$h_{30}$ & $h_{40}$& $h_{35}$& $h_{45}$& $y_{30}$&$y_{40}$ & $y_{35}$& $y_{45} $ \\
\hline
  9.22582 & 58.5642 & 37.392 &8.26747 &1.0 & 6.0  & 8.0 & 10.0   \\
\hline
\end{tabular}
\caption{The set of coupling constants that we have used to study
the effect of unequal $y$ values and large $h$ values.
\label{tb:coupling2}}
\end{table}
 
\section{Conclusions and outlook}
\label{concl}
In this paper, we have studied the strong Yukawa dynamics of the mirror sector
of the anomaly-free 3-4-5 model.  We studied the polarization operator of the mirror 
theory at zero gauge field background, at values of the mirror couplings where the 
fluctuations in the phase of the mirror-fermion determinant are small, in order to probe 
for the existence of massless mirror states. Massless states are expected to manifest 
themselves via a zero-momentum directional discontinuity (a characteristic of the 
two dimensional case). 

Multiple checks of the rather lengthy code (see the expression for $\Pi'_{\mu\nu}$ in 
Appendix \ref{exppi}) were performed, using several coupling-independent properties that
must be satisfied by any mirror polarization operator. These properties are obeyed up to 
the errors of the simulation, see Tables~\ref{transcheck1}, \ref{transcheck2}, \ref{antitab},
and give us confidence in the results that we now summarize.

Our main result is shown on Figure~\ref{fig:345discfit}: we found that as we increase the 
volume of the lattice $N=L/a$ in lattice units, which is equivalent
to taking the continuum limit $a \to 0$ with $L$ held fixed, 
the discontinuity in the polarization tensor remains essentially constant
and has a nonzero value in the continuum limit.  We thus have evidence that in the 
continuum limit the polarization tensor has a directional 
discontinuity at $k=0$; our study also suggests that 
there is at best a weak coupling-constant dependence of the analytic behavior 
of $\Pi_{\mu\nu}$ at $k=0$.  We compared the behavior of the discontinuity 
in the 3-4-5 model to that for a free massless charged Weyl Ginsparg-Wilson fermion, 
computed exactly, on the same size lattices, and shown on Figure~\ref{fig:freemasslessdisc}. 
Clearly, the behavior is quite similar to that found in the 3-4-5 model.   
Comparing to the case of a massive Ginsparg-Wilson fermion, Fig.~\ref{fig:massdiscfit}, we find that
the 3-4-5 model is quite different from that picture, supporting the conclusion that the mirror sector
fermions do not decouple, but remain massless, in spite of the strong Yukawa dynamics.
As discussed in the end of Section \ref{latsize}, the results for the discontinuity are consistent with either a chiral massless mirror fermion spectrum (with mirror fermions of charges and chirality $3_-$, $4_-$, $5_+$ remaining massless) or a vectorlike spectrum (with massless mirror fermions of charges and chirality $5_-$, $5_+$).
 
 We end with some comments and discussion of possible future work.
 
The study of this paper was motivated by  the fact that 't 
Hooft anomaly matching  does not demand that there be massless states in the anomaly-free 
case, provided all global mirror chiral symmetries are explicitly broken. The use of lattice fermions with exact chiral symmetries allows a precise formulation of  anomaly matching on the lattice. Furthermore, exact lattice chirality helps construct potential mirror-decoupling  models where all global chiral symmetries in the mirror sector are explicitly broken, while the chiral symmetries protecting the light fermions are manifest. Such a clear symmetry realization was not possible before the advent of exact lattice chirality (we note that earlier Ref.~\cite{Creutz:1996xc} proposed a lattice formulation of the Standard Model similar  to Eichten-Preskill and to the setup studied here, but in the domain wall framework). 
  However, anomaly matching alone does  
not forbid massless states in anomaly-free representations (these would be accompanied by the appearance of ``emergent" global chiral symmetries). 
The general 
 belief that massless states always exist ``for a reason" gives rise to the expectation that  a mirror sector  breaking all non-gauged global symmetries   ``has no reason to" and therefore will not exhibit massless states. Yet, this expectation must be checked in practice, especially  when faced with a poorly understood strong mirror dynamics where there might be  a yet-to-be-found dynamical reason for
 massless states to exist at any coupling.  
 
 Indeed, the evidence presented in this paper
suggests that this is the case in the 3-4-5 model that we have studied.
  The appearance of  massless mirror fermions  in an anomaly-free representation---chiral ($3_-$, $4_-$, $5_+$) or vectorlike ($5_+$, $5_-$)---at strong mirror couplings implies that there are ``emergent" (i.e., infrared) chiral global symmetries,  appearing despite the fact that  the mirror action explicitly breaks all mirror global chiral symmetries. We think that it would be of   interest to better understand the reason why this enhancement of chiral symmetries occurs: to satisfy curiosity and maybe finally close the door to the ``mirror decoupling" approach or, perhaps, to further modify the idea  to prevent the emergence of  chiral symmetries. At the moment, it is not clear what form, if any, such modifications of ``mirror decoupling" might take---perhaps additional interactions between the mirror fermions, or understanding the role of gauge interactions.
  
  In this regard, it may be useful to recall the studies of the phases of ``waveguide" models at strong Yukawa coupling  \cite{Golterman:1992yha,
  Golterman:1994at}, where in the strong coupling limit a large number of doublers appears at the waveguide boundary (supposedly containing the decoupled mirror fermions), rendering the spectrum vectorlike. A better understanding of the connection between these models and the studies of decoupling via lattice fermions with exact chiral symmetries might be useful. 
  The major difference between the setups studied in these references and the present study is the realization of exact chiral symmetries at finite lattice spacing. The relation between the finite-size domain wall  and the overlap has been understood in great detail  \cite{Kikukawa:1999sy}, see \cite{Fosco:2007ry} for a more heuristic derivation, in the absence of Yukawa couplings, but not when they are present, especially when taken to be strong; see the comments in Section 5 of \cite{Poppitz:2009gt}. The results found here provide us with some  motivation to return to this problem. 
 
\acknowledgments

C.C.~and J.G.~were supported in part by the Department of Energy,
Office of Science, Office of High Energy Physics, 
Grant No.~DE-FG02-08ER41575.  
E.P.~was supported in part by the National Science 
and Engineering Council of Canada (NSERC).
We gratefully acknowledge
the use of USQCD computing resources at Fermi National
Laboratory, under a Class C allocation.

\appendix

\clearpage
\pagebreak

\section{Expression of $\Pi_{\mu\nu}$}
\label{exppi} 
\beq \label{eq:exp_PI}
\Pi_{\mu \nu} &=& \delta_\nu ( j_\mu^{wA} + j_\mu^{wB} + j_\mu^{uC})
\ddd + \vev{ \bar \alpha_{-}^i \alpha_{-}^j } 
(w_{iA}^\dagger \cdot ( \delta_\nu \delta_\mu D_3 + \delta_\nu \hat P_{+A} 
\cdot \delta_\mu D_3 ) \cdot t_{j})
\ddd + \vev{ \bar \beta_{-}^i \beta_{-}^j } (w_{iB}^\dagger \cdot (\delta_\nu \delta_\mu D_4 
+ \delta_\nu \hat P_{+B} \cdot \delta_\mu D_4) \cdot t_{j})
\ddd + \vev{ \bar \gamma_{+}^i \gamma_{+}^j } (u_{iC}^\dagger \cdot (\delta_\nu \delta_\mu D_5 
+ \delta_\nu \hat P_{-C} \cdot \delta_\mu D_5 ) \cdot v_{j})
\ddd
+ \bigg\langle [ \bar \alpha_{-}^i \alpha_{-}^j (w_{iA}^\dagger \cdot \delta_\mu D_3 \cdot t_{j})
+ \bar \beta_{-}^i \beta_{-}^j (w_{iB}^\dagger \cdot \delta_\mu D_4 \cdot t_{j})
\ddd + \bar \gamma_{+}^i \gamma_{+}^j (u_{iC}^\dagger \cdot \delta_\mu D_5 \cdot v_{j}) ]
[ \bar \alpha_{-}^k \alpha_{-}^l (w_{kA}^\dagger \cdot \delta_\nu D_3 \cdot t_{l})
\ddd + \bar \beta_{-}^k \beta_{-}^l (w_{kB}^\dagger \cdot \delta_\nu D_4 \cdot t_{l})
+ \bar \gamma_{+}^k \gamma_{+}^l (u_{kC}^\dagger \cdot \delta_\nu D_5 \cdot v_{l}) ] \bigg\rangle^C
\ddd + \frac{\kappa}{2} \vev{ (\phi^* \cdot \delta_\nu \delta_\mu U^* \cdot \phi) + \hc }
\ddd + \frac{\kappa^2}{4} \vev{ [(\phi^* \cdot \delta_\mu U^* \cdot \phi) + \hc]
[(\phi^* \cdot \delta_\nu U^* \cdot \phi) + \hc] }^C
\ddd + \frac{\kappa}{2} \bigg\langle [(\phi^* \cdot \delta_\mu U^* \cdot \phi) + \hc] 
[ \bar \alpha_{-}^i \alpha_{-}^j (w_{iA}^\dagger \cdot \delta_\mu D_3 \cdot t_{j})
+ \bar \beta_{-}^i \beta_{-}^j (w_{iB}^\dagger \cdot \delta_\mu D_4 \cdot t_{j})
\ddd + \bar \gamma_{+}^i \gamma_{+}^j (u_{iC}^\dagger \cdot \delta_\mu D_5 \cdot v_{j}) ] 
+ (\mu \leftrightarrow \nu) \bigg\rangle^C
\ddd - y_{30} \vev{\bar\alpha_-^i \chi_+^j }(w_{iA}^\dagger 
\cdot \delta_\nu (\hat P_{+A} \cdot \delta_\mu \hat P_{+A}) \cdot \phi^{-3} \cdot v_j)
- y_{40} \vev{\bar\beta_-^i \chi_+^j} (w_{iB}^\dagger 
\cdot \delta_\nu (\hat P_{+B} \cdot \delta_\mu \hat P_{+B}) \cdot \phi^{-4} \cdot v_j)
\ddd - y_{35} \vev{ \bar\alpha_-^i \gamma_+^j} (w_{iA}^\dagger 
\cdot \delta_\nu (\hat P_{+A} \cdot \delta_\mu \hat P_{+A}) \cdot \phi^{2} \cdot v_j)
- y_{45} \vev{\bar\beta_-^i \gamma_+^j} (w_{iB}^\dagger \cdot 
\delta_\nu (\hat P_{+B} \cdot \delta_\mu \hat P_{+B}) \cdot \phi \cdot v_j)
\ddd - y_{35} \vev{\bar\gamma_+^i \alpha_-^j} (u_{iC}^\dagger \cdot 
\delta_\nu (\hat P_{-C} \cdot \delta_\mu \hat P_{-C}) \cdot \phi^{-2} \cdot t_j)
- y_{45} \vev{\bar\gamma_+^i \beta_-^j} (u_{iC}^\dagger \cdot 
\delta_\nu (\hat P_{-C} \cdot \delta_\mu \hat P_{-C}) \cdot \phi^{-1} \cdot t_j)
\ddd + h_{30} \vev{ \bar\chi_+^i \bar\alpha_-^j } ( u_{iX}^\dagger \cdot \phi^{-3} 
\cdot \gamma_2 \cdot \delta_\nu (\delta_\mu \hat P_{+A}^T \cdot \hat P_{+A}^T) \cdot w_{jA}^* ) 
\ddd + h_{40} \vev{ \bar\chi_+^i \bar\beta_-^j } (u_{iX}^\dagger \cdot \phi^{-4} 
\cdot \gamma_2 \cdot \delta_\nu (\delta_\mu \hat P_{+B}^T \cdot \hat P_{+B}^T) \cdot w_{jB}^* ) 
\ddd + h_{35} \vev{ \bar\gamma_+^i \bar\alpha_-^j } [ (u_{iC}^\dagger 
\cdot \delta_\nu (\hat P_{-C} \cdot \delta_\mu \hat P_{-C}) \cdot \phi^{-8} 
\cdot \gamma_2 \cdot w_{jA}^* ) \ddd + (u_{iC}^\dagger \cdot \phi^{-8} 
\cdot \gamma_2 \cdot \delta_\nu (\delta_\mu \hat P_{+A}^T \cdot \hat P_{+A}^T) \cdot w_{jA}^* ) ]
\ddd + h_{35} \vev{ \bar\gamma_+^i \bar\alpha_-^j } [ (u_{iC}^\dagger \cdot \delta_\mu \hat P_{-C}
\cdot \phi^{-8} \cdot \gamma_2 \cdot \delta_\nu \hat P_{+A}^T \cdot w_{jA}^* ) 
+ (\mu \leftrightarrow \nu) ]
\ddd + h_{45} \vev{ \bar\gamma_+^i \bar\beta_-^j } [ (u_{iC}^\dagger \cdot \delta_\mu \hat P_{-C}
\cdot \phi^{-9} \cdot \gamma_2 \cdot \delta_\nu \hat P_{+B}^T \cdot w_{jB}^* ) 
+ (\mu \leftrightarrow \nu) ]
\ddd + h_{45} \vev{ \bar\gamma_+^i \bar\beta_-^j } [ (u_{iC}^\dagger 
\cdot \delta_\nu (\hat P_{-C} \cdot \delta_\mu \hat P_{-C}) \cdot \phi^{-9} 
\cdot \gamma_2 \cdot w_{jB}^* ) \ddd + (u_{iC}^\dagger \cdot \phi^{-9} 
\cdot \gamma_2 \cdot \delta_\nu (\delta_\mu \hat P_{+B}^T \cdot \hat P_{+B}^T) \cdot w_{jB}^* ) ]
\ddd  \text{(continued)} \nonumber
\eeq
\beq
\quad && - \bigg\langle \big[ \bar \alpha_{-}^k \alpha_{-}^l (w_{kA}^\dagger 
\cdot \delta_\mu D_3 \cdot t_{l})
+ \bar \beta_{-}^k \beta_{-}^l (w_{kB}^\dagger \cdot \delta_\mu D_4 \cdot t_{l})
\ddd + \bar \gamma_{+}^k \gamma_{+}^l (u_{kC}^\dagger \cdot \delta_\mu D_5 \cdot v_{l}) \big]
\ddd \times \big[ y_{30} \bar\alpha_-^i \chi_+^j (w_{iA}^\dagger \cdot \delta_\nu 
\hat P_{+A} \cdot \phi^{-3} \cdot v_j)
+ y_{40} \bar\beta_-^i \chi_+^j (w_{iB}^\dagger \cdot \delta_\nu \hat P_{+B} \cdot \phi^{-4} \cdot v_j)
\ddd + y_{35} \bar\alpha_-^i \gamma_+^j (w_{iA}^\dagger \cdot \delta_\nu 
\hat P_{+A} \cdot \phi^{2} \cdot v_j)
+ y_{45} \bar\beta_-^i \gamma_+^j (w_{iB}^\dagger \cdot \delta_\nu \hat P_{+B} \cdot \phi \cdot v_j)
\ddd + y_{35} \bar\gamma_+^i \alpha_-^j (u_{iC}^\dagger \cdot \delta_\nu 
\hat P_{-C} \cdot \phi^{-2} \cdot t_j)
+ y_{45} \bar\gamma_+^i \beta_-^j (u_{iC}^\dagger \cdot \delta_\nu \hat P_{-C} \cdot \phi^{-1} \cdot t_j) 
\ddd - h_{30} \bar\chi_+^i \bar\alpha_-^j (u_{iX}^\dagger \cdot \phi^{-3} 
\cdot \gamma_2 \cdot \delta_\nu \hat P_{+A}^T \cdot w_{jA}^* ) 
- h_{40} \bar\chi_+^i \bar\beta_-^j (u_{iX}^\dagger \cdot \phi^{-4} 
\cdot \gamma_2 \cdot \delta_\nu \hat P_{+B}^T \cdot w_{jB}^* ) 
\ddd - h_{35} \bar\gamma_+^i \bar\alpha_-^j [(u_{iC}^\dagger \cdot \delta_\nu \hat P_{-C} \cdot \phi^{-8} 
\cdot \gamma_2 \cdot w_{jA}^* )+(u_{iC}^\dagger \cdot \phi^{-8} \cdot  
\gamma_2 \cdot \delta_\nu \hat P_{+A}^T \cdot w_{jA}^* )]
\ddd - h_{45} \bar\gamma_+^i \bar\beta_-^j [(u_{iC}^\dagger \cdot \delta_\nu \hat P_{-C} \cdot \phi^{-9} 
\cdot \gamma_2 \cdot w_{jB}^* )
\ddd +(u_{iC}^\dagger \cdot \phi^{-9} 
\cdot \gamma_2 \cdot \delta_\nu \hat P_{+B}^T \cdot w_{jB}^* )]
\big] + (\mu \leftrightarrow \nu) \bigg\rangle^C
\ddd + \bigg\langle 
\big\{ y_{30} \bar\alpha_-^i \chi_+^j 
(w_{iA}^\dagger \cdot \delta_\mu \hat P_{+A} \cdot \phi^{-3} \cdot v_j)
+ y_{40} \bar\beta_-^i \chi_+^j (w_{iB}^\dagger \cdot \delta_\mu \hat P_{+B} \cdot \phi^{-4} \cdot v_j)
\ddd + y_{35} \bar\alpha_-^i \gamma_+^j (w_{iA}^\dagger \cdot \delta_\mu 
\hat P_{+A} \cdot \phi^{2} \cdot v_j)
+ y_{45} \bar\beta_-^i \gamma_+^j (w_{iB}^\dagger \cdot \delta_\mu \hat P_{+B} \cdot \phi \cdot v_j)
\ddd + y_{35} \bar\gamma_+^i \alpha_-^j (u_{iC}^\dagger \cdot \delta_\mu 
\hat P_{-C} \cdot \phi^{-2} \cdot t_j)
+ y_{45} \bar\gamma_+^i \beta_-^j (u_{iC}^\dagger \cdot \delta_\mu \hat P_{-C} \cdot \phi^{-1} \cdot t_j) 
\ddd -h_{30} \bar\chi_+^i \bar\alpha_-^j (u_{iX}^\dagger \cdot \phi^{-3} 
\cdot \gamma_2 \cdot \delta_\mu \hat P_{+A}^T \cdot w_{jA}^* ) 
-h_{40} \bar\chi_+^i \bar\beta_-^j (u_{iX}^\dagger \cdot \phi^{-4} 
\cdot \gamma_2 \cdot \delta_\mu \hat P_{+B}^T \cdot w_{jB}^* ) 
\ddd -h_{35} \bar\gamma_+^i \bar\alpha_-^j [ (u_{iC}^\dagger \cdot \delta_\mu \hat P_{-C} \cdot \phi^{-8} 
\cdot \gamma_2 \cdot w_{jA}^* ) + (u_{iC}^\dagger \cdot \phi^{-8} 
\cdot \gamma_2 \cdot \delta_\mu \hat P_{+A}^T \cdot w_{jA}^* ) ]
\ddd -h_{45} \bar\gamma_+^i \bar\beta_-^j [(u_{iC}^\dagger \cdot \delta_\mu \hat P_{-C} \cdot \phi^{-9} 
\cdot \gamma_2 \cdot w_{jB}^* ) + (u_{iC}^\dagger \cdot \phi^{-9} 
\cdot \gamma_2 \cdot \delta_\mu \hat P_{+B}^T \cdot w_{jB}^* )]
\big\} 
\ddd \times \big\{ y_{30} \bar\alpha_-^k \chi_+^l (w_{kA}^\dagger \cdot \delta_\nu 
\hat P_{+A} \cdot \phi^{-3} \cdot v_l)
+ y_{40} \bar\beta_-^k \chi_+^l (w_{kB}^\dagger \cdot \delta_\nu \hat P_{+B} \cdot \phi^{-4} \cdot v_l)
\ddd + y_{35} \bar\alpha_-^k \gamma_+^l (w_{kA}^\dagger \cdot \delta_\mu 
\hat P_{+A} \cdot \phi^{2} \cdot v_l)
+ y_{45} \bar\beta_-^k \gamma_+^l (w_{kB}^\dagger \cdot \delta_\nu \hat P_{+B} \cdot \phi \cdot v_l)
\ddd + y_{35} \bar\gamma_+^k \alpha_-^l (u_{kC}^\dagger \cdot \delta_\nu 
\hat P_{-C} \cdot \phi^{-2} \cdot t_l)
+ y_{45} \bar\gamma_+^k \beta_-^l (u_{kC}^\dagger \cdot \delta_\nu \hat P_{-C} \cdot \phi^{-1} \cdot t_l) 
\ddd -h_{30} \bar\chi_+^k \bar\alpha_-^l (u_{kX}^\dagger \cdot \phi^{-3} 
\cdot \gamma_2 \cdot \delta_\nu \hat P_{+A}^T \cdot w_{lA}^* ) 
-h_{40} \bar\chi_+^k \bar\beta_-^l (u_{kX}^\dagger \cdot \phi^{-4} 
\cdot \gamma_2 \cdot \delta_\nu \hat P_{+B}^T \cdot w_{lB}^* ) 
\ddd - h_{35} \bar\gamma_+^k \bar\alpha_-^l [ (u_{kC}^\dagger \cdot \delta_\nu \hat P_{-C} \cdot \phi^{-8} 
\cdot \gamma_2 \cdot w_{lA}^* ) + (u_{kC}^\dagger \cdot \phi^{-8} 
\cdot \gamma_2 \cdot \delta_\nu \hat P_{+A}^T \cdot w_{lA}^* ) ]
\ddd -h_{45} \bar\gamma_+^k \bar\beta_-^l [(u_{kC}^\dagger \cdot \delta_\nu \hat P_{-C} \cdot \phi^{-9} 
\cdot \gamma_2 \cdot w_{lB}^* ) + (u_{kC}^\dagger \cdot \phi^{-9} 
\cdot \gamma_2 \cdot \delta_\nu \hat P_{+B}^T \cdot w_{lB}^* )]
\big\} \bigg\rangle^C
\ddd \text{(continued)} \nonumber
\eeq
\beq
\quad && - \frac{\kappa}{2} \bigg\langle [(\phi^* \cdot \delta_\mu U^* \cdot \phi) + \hc]
\ddd \times \big\{ y_{30} \bar\alpha_-^i \chi_+^j 
(w_{iA}^\dagger \cdot \delta_\nu \hat P_{+A} \cdot \phi^{-3} \cdot v_j)
+ y_{40} \bar\beta_-^i \chi_+^j (w_{iB}^\dagger \cdot \delta_\nu \hat P_{+B} \cdot \phi^{-4} \cdot v_j)
\ddd + y_{35} \bar\alpha_-^i \gamma_+^j (w_{iA}^\dagger \cdot \delta_\nu 
\hat P_{+A} \cdot \phi^{2} \cdot v_j)
+ y_{45} \bar\beta_-^i \gamma_+^j (w_{iB}^\dagger \cdot \delta_\nu \hat P_{+B} \cdot \phi \cdot v_j)
\ddd + y_{35} \bar\gamma_+^i \alpha_-^j (u_{iC}^\dagger \cdot \delta_\nu 
\hat P_{-C} \cdot \phi^{-2} \cdot t_j)
+ y_{45} \bar\gamma_+^i \beta_-^j (u_{iC}^\dagger \cdot \delta_\nu \hat P_{-C} \cdot \phi^{-1} \cdot t_j) 
\ddd -h_{30} \bar\chi_+^i \bar\alpha_-^j (u_{iX}^\dagger \cdot \phi^{-3} 
\cdot \gamma_2 \cdot \delta_\nu \hat P_{+A}^T \cdot w_{jA}^* ) 
-h_{40} \bar\chi_+^i \bar\beta_-^j (u_{iX}^\dagger \cdot \phi^{-4} 
\cdot \gamma_2 \cdot \delta_\nu \hat P_{+B}^T \cdot w_{jB}^* ) 
\ddd -h_{35} \bar\gamma_+^i \bar\alpha_-^j [ (u_{iC}^\dagger \cdot \delta_\nu \hat P_{-C} \cdot \phi^{-8} 
\cdot \gamma_2 \cdot w_{jA}^* ) + (u_{iC}^\dagger \cdot \phi^{-8} 
\cdot \gamma_2 \cdot \delta_\nu \hat P_{+A}^T \cdot w_{jA}^* ) ]
\ddd -h_{45} \bar\gamma_+^i \bar\beta_-^j [(u_{iC}^\dagger \cdot \delta_\nu \hat P_{-C} \cdot \phi^{-9} 
\cdot \gamma_2 \cdot w_{jB}^* ) 
\ddd + (u_{iC}^\dagger \cdot \phi^{-9} 
\cdot \gamma_2 \cdot \delta_\nu \hat P_{+B}^T \cdot w_{jB}^* )]
\big\} + (\mu \leftrightarrow \nu) \bigg\rangle^C.
\eeq 

\section{Fermion Matrix}
\label{fmat}
Our purpose here is to rewrite the mirror fermion action
after the change of variables \myref{eqme}.
In the limit of large Yukawa couplings, where we neglect the
kinetic terms,
\beq
S_{\text{mirror, fermions}} =
(\alphabar_k , \alpha_k, \betabar_k , \beta_k )
{\cal M}_{kp}
\begin{pmatrix}
\chi_p \cr \chib_p \cr
\gamma_p \cr \gammab_p
\end{pmatrix}
\eeq
The fermion matrix ${\cal M}$ can be expressed in the block diagonal form
\beq
{\cal M}_{kp} =
\begin{pmatrix}
M_{kp}^{(3,0)} & M_{kp}^{(3,5)} \cr
M_{kp}^{(4,0)} & M_{kp}^{(4,5)}
\end{pmatrix}
\eeq
Using notations given in \cite{Giedt:2007qg}, the
blocks are equal to
\beq
(M_{kp}^{(3,0)})_{1,1} &=& \half y_{30} (2-\lambda_k) \Phi_{k-p}^{(-3)} \nnn
(M_{kp}^{(3,0)})_{2,2} &=& -\half y_{30} (2-\lambda_p) 
e^{i(\varphi_k-\varphi_p)} \Phi_{p-k}^{(3)} \nnn
(M_{kp}^{(3,0)})_{1,2} &=& \frac{ih_{30}}{4} \[ (2-\lambda_k)(2-\lambda_p)e^{-i\varphi_p}
- \lambda_p \lambda_k e^{-i\varphi_k} \] \Phi_{k+p}^{(-3)} \nnn
(M_{kp}^{(3,0)})_{2,1} &=& ih_{30} e^{i\varphi_k} \Phi_{-p-k}^{(3)}
\eeq

\beq
(M_{kp}^{(4,0)})_{1,1} &=& \half y_{40} (2-\lambda_k) \Phi_{k-p}^{(-4)} \nnn
(M_{kp}^{(4,0)})_{2,2} &=& -\half y_{40} (2-\lambda_p) 
e^{i(\varphi_k-\varphi_p)} \Phi_{p-k}^{(4)} \nnn
(M_{kp}^{(4,0)})_{1,2} &=& \frac{ih_{40}}{4} \[ (2-\lambda_k)(2-\lambda_p)e^{-i\varphi_p}
- \lambda_p \lambda_k e^{-i\varphi_k} \] \Phi_{k+p}^{(-4)} \nnn
(M_{kp}^{(4,0)})_{2,1} &=& ih_{40} e^{i\varphi_k} \Phi_{-p-k}^{(4)}
\eeq

\beq
(M_{kp}^{(3,5)})_{1,1} &=& \half y_{35} (2-\lambda_k) \Phi_{k-p}^{(2)} \nnn
(M_{kp}^{(3,5)})_{2,2} &=& -\half y_{35} (2-\lambda_p) 
e^{i(\varphi_k-\varphi_p)} \Phi_{p-k}^{(-2)} \nnn
(M_{kp}^{(3,5)})_{1,2} &=& \frac{ih_{35}}{4} \[ (2-\lambda_k)(2-\lambda_p)e^{-i\varphi_p}
- \lambda_p \lambda_k e^{-i\varphi_k} \] \Phi_{k+p}^{(-8)} \nnn
(M_{kp}^{(3,5)})_{2,1} &=& ih_{35} e^{i\varphi_k} \Phi_{-p-k}^{(8)}
\eeq

\beq
(M_{kp}^{(4,5)})_{1,1} &=& \half y_{45} (2-\lambda_k) \Phi_{k-p}^{(1)} \nnn
(M_{kp}^{(4,5)})_{2,2} &=& -\half y_{45} (2-\lambda_p) 
e^{i(\varphi_k-\varphi_p)} \Phi_{p-k}^{(-1)} \nnn
(M_{kp}^{(4,5)})_{1,2} &=& \frac{ih_{45}}{4} \[ (2-\lambda_k)(2-\lambda_p)e^{-i\varphi_p}
- \lambda_p \lambda_k e^{-i\varphi_k} \] \Phi_{k+p}^{(-9)} \nnn
(M_{kp}^{(4,5)})_{2,1} &=& ih_{45} e^{i\varphi_k} \Phi_{-p-k}^{(9)}
\eeq

\bibliography{gw345_rev}
\bibliographystyle{JHEP}

\end{document}